\documentclass[%
  reprint,
  superscriptaddress,
  showpacs,
  amsmath,
  amssymb,
  aps,
  prl,
  floatfix,
]{revtex4-2}

\usepackage{xcolor}
\usepackage{marginnote}

\usepackage{graphicx}
\usepackage{dcolumn}
\usepackage{bm}
\usepackage{enumerate}

\usepackage{hyperref}

\usepackage{orcidlink}

\begin{document}

\preprint{APS/123-QED}

\title{High-Fidelity Individual Addressing of Single Atoms in Quantum Registers at Three-Photon Laser Excitation of Rydberg States} 

\author{
  N.N.Bezuglov\,\orcidlink{0000-0003-0191-988X}
}
\affiliation{Saint-Petersburg State University, 199034 St.Petersburg, Russia}
\affiliation{Rzhanov Institute of Semiconductor Physics SB RAS, 630090 Novosibirsk, Russia}

\author{
  I.I.Beterov\,\orcidlink{0000-0002-6596-6741}
}
\affiliation{Rzhanov Institute of Semiconductor Physics SB RAS, 630090 Novosibirsk, Russia}
\affiliation{Novosibirsk State University, Department of Physics, 630090 Novosibirsk, Russia}

\author{
  A.Cinins\,\orcidlink{0000-0002-3660-2033}
}
\email[]{arturs.cinins@lu.lv}
\affiliation{Institute of Atomic Physics and Spectroscopy, Faculty of Science and Technology, University of Latvia, Jelgavas Street 3, LV-1004 Riga, Latvia}

\author{
  K.Miculis\,\orcidlink{0009-0009-5635-9817}
}
\affiliation{Institute of Atomic Physics and Spectroscopy, Faculty of Science and Technology, University of Latvia, Jelgavas Street 3, LV-1004 Riga, Latvia}
\affiliation{National Research Nuclear University MEPhI, Moscow, 115409 Russia}

\author{
  V.M.Entin\,\orcidlink{0000-0001-5436-2849}
}
\affiliation{Rzhanov Institute of Semiconductor Physics SB RAS, 630090 Novosibirsk, Russia}

\author{
  P.I.Betleni\,\orcidlink{0009-0006-2964-6229}
}
\affiliation{Rzhanov Institute of Semiconductor Physics SB RAS, 630090 Novosibirsk, Russia}
\affiliation{Novosibirsk State University, Department of Physics, 630090 Novosibirsk, Russia}

\author{
  G.Suliman\,\orcidlink{0009-0000-3769-4235}
}
\affiliation{Rzhanov Institute of Semiconductor Physics SB RAS, 630090 Novosibirsk, Russia}
\affiliation{Novosibirsk State University, Department of Physics, 630090 Novosibirsk, Russia}

\author{
  V.V.Gromyko\,\orcidlink{0009-0004-5026-3535}
}
\affiliation{Rzhanov Institute of Semiconductor Physics SB RAS, 630090 Novosibirsk, Russia}

\author{
  D.B.Tretyakov\,\orcidlink{0000-0002-3708-6253}
}
\affiliation{Rzhanov Institute of Semiconductor Physics SB RAS, 630090 Novosibirsk, Russia}
\affiliation{Novosibirsk State University, Department of Physics, 630090 Novosibirsk, Russia}

\author{
  E.A.Yakshina\,\orcidlink{0009-0000-7621-2454}
}
\affiliation{Rzhanov Institute of Semiconductor Physics SB RAS, 630090 Novosibirsk, Russia}
\affiliation{Novosibirsk State University, Department of Physics, 630090 Novosibirsk, Russia}

\author{
  I.I.Ryabtsev\,\orcidlink{0000-0002-5410-2155}
}
\affiliation{Rzhanov Institute of Semiconductor Physics SB RAS, 630090 Novosibirsk, Russia}
\affiliation{Novosibirsk State University, Department of Physics, 630090 Novosibirsk, Russia}

\date{\today}

\begin{abstract}
Precise individual addressing of single atoms in quantum registers formed by optical trap arrays is essential to achieve high-fidelity quantum gates in neutral-atom quantum computers and simulators. 
Two-qubit quantum gates are typically realized using coherent two-photon laser excitation of atoms to strongly interacting Rydberg states. 
However, two-photon excitation encounters challenges in individual addressing with tightly focused laser beams due to atom position uncertainty and the spatial inhomogeneity in both Rabi frequencies and light shifts. 
In this work, we theoretically demonstrate that the fidelity of individual addressing can be improved by employing coherent three-photon laser excitation of Rydberg states.
For a specific example of $5s_{1/2}\!\xrightarrow{\Omega_1}\!5p_{3/2}\!\xrightarrow{\Omega_2}\!6s_{1/2}\!\xrightarrow{\Omega_3}\!np$ excitation in $^{87}$Rb atoms, we find that upon strong laser coupling in the second step (Rabi frequency $\Omega_2$) and moderate coupling in the first and third steps (Rabi frequencies $\Omega_1$ and $\Omega_3$), the three-photon Rabi frequency is given by $\Omega \!=\!\Omega_1\Omega_3/\Omega_2$.
If the spatial distributions of $(\Omega_1\Omega_3)$ and $\Omega_2$ are arranged to be identical, $\Omega$ becomes independent of atom position, even within very tightly focused laser beams. 
This approach dramatically improves individual addressing of Rydberg excitation for neighboring atoms in trap arrays compared to conventional two-photon excitation schemes.
Our findings are crucial for large-scale quantum registers of neutral atoms, where distances between adjacent atoms should be minimized to ensure stronger Rydberg interactions and compact arrangement of atom arrays.
\end{abstract}



\maketitle

Neutral-atom-based quantum computers and simulators represent a rapidly growing research field, which is close to practical realization \cite{Saffman2010,Saffman2016,Ryabtsev2016,Henriet2020,Shi2022}.
The main advantage of this platform is its nearly unlimited potential for scalability to very large numbers of qubits, represented by single atoms trapped in optical dipole trap arrays \cite{Barredo2016}.
Several recent studies have demonstrated quantum registers with thousands of qubits \cite{Pause2024,Pichard2024,manetsch2024}.
High-fidelity single-qubit gates in atomic arrays have been demonstrated both without individual addressing \cite{Sheng2018,Nikolov2023} and with individual addressing \cite{Ma2023,Jenkins2022}, though with a slight reduction in fidelity.
Implementing two-qubit gates, which are essential for a universal quantum computer, poses a greater challenge.
These gates rely on laser excitation to strongly interacting Rydberg states and on the associated Rydberg blockade effect \cite{Saffman2010,Saffman2016,Ryabtsev2016,Henriet2020,Shi2022,Lukin2001}.
High-fidelity two-qubit gates using spatially homogeneous laser beams for Rydberg excitation have recently been demonstrated~\cite{Levine2018,Levine2019,Ebadi2021,Fu2022,Evered2023,Bluvstein2023,Radnaev2024,Tsai2024}, with fidelities reaching $0.997$~\cite{Tsai2024}.
These fidelities are comparable to those achieved in superconducting~\cite{Huang2020} or ion~\cite{Bruzewicz2019} qubits, which, however, have not yet been scaled to thousands of qubits.
Unfortunately, using tightly focused laser beams for individual addressing of single pairs of atoms in the array reduces two-qubit gate fidelities to around 0.95~\cite{Graham2022}.
This presents a significant obstacle to building a universal quantum computer.
Without individual addressing, coherent transport in atomic arrays partially resolves this issue by controlling interatomic interaction energy through adjustments in atom spacing~\cite{Ebadi2021}.
However, this approach requires constant rearrangement of the atomic array for each quantum algorithm, which is technically challenging.

The primary challenge in achieving higher fidelity with neutral atoms is the finite atom temperature, typically in the microkelvin range for optical dipole trap experiments.
While lower temperatures are possible through sideband Raman cooling \cite{Tian2024}, laser excitation to Rydberg states reheats the atoms to the microkelvin scale.
Residual thermal atom motion introduces inhomogeneity in the Rabi frequency of Rydberg excitation when tightly focused laser beams are used, as the local laser field varies with atom position within the beam.
In two-photon excitation schemes, spatially inhomogeneous light shifts that depend on atom position also induce decoherence in Rydberg excitation.
To reduce inhomogeneity effects in two-photon excitation, less focused laser beams are required, which reduces single-qubit addressability and increases cross-talk with neighboring qubits.
Mitigating the cross-talk requires increased spacing between adjacent atoms, yet in large-scale quantum registers, minimizing atom distances is essential to ensure stronger Rydberg interactions and a compact atomic array layout.

\begin{figure}
    \includegraphics[width=\linewidth]{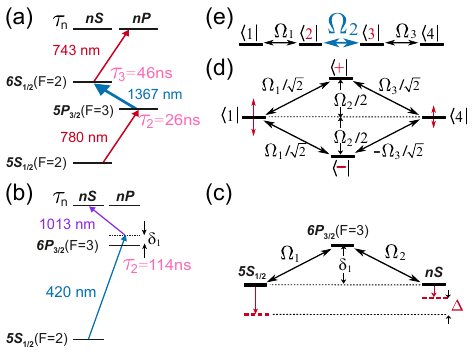}
    \caption{
      (a) Three-photon \cite{Ryabtsev2016,Entin2013,Yakshina2018,Yakshina2020,Tretyakov2022,Beterov2023,Beterov2024} and (b) two-photon \cite{Levine2018,Levine2019,Ebadi2021,Evered2023,Bluvstein2023} excitation schemes of Rydberg $nP$, $nS$ states in $^{87}$Rb atoms with the corresponding linkage diagrams (e,d), (c) in the rotating-wave approximation. 
      The case of exact resonances [$\delta _{1,2,3}\!=\!0$ in Eq.~(3)] for all three lasers in frames (d,e) and two-photon resonance with single-photon detuning $\delta_1$ in frame (c) are presented. 
      All intermediate states, with lifetimes $\tau$ marked in pink, become virtual either due to the high intensity ($\Omega _2\!\gg\! \Omega _{1,3}$) of the intermediate laser [frames (e,d)] or due to the large ($\delta _1 \!\gg\! \Omega _{1,2}$) single-photon detuning [frame (c)]. 
      Both the ground $5S _{1/2}$ and the Rydberg $nS$ states in frame (c) experience different ac Stark shifts (marked in red), leading to a spatially inhomogeneous two-photon detuning $\Delta (r)$~(\ref{eq:2Rabi}). 
      In frame (e), a very strong laser coupling between the intermediate states $|2\rangle, |3\rangle$ transforms them into virtual $|\pm \rangle \!= \!(|2 \rangle \!\pm \!|3 \rangle)/\sqrt{2}$ states as shown in frame (d). 
      Due to the resulting mirror-symmetric atomic states configuration, each of the states $|1\rangle, |4\rangle$ experiences two identical ac Stark shifts (indicated by the red vertical arrows) with opposite signs, which makes the aggregate light shifts of states $|1 \rangle ,|4 \rangle$ vanish for any choice of Rabi frequencies $\Omega _{1},\Omega _{3}$. 
    }
    \label{fig:Lukin}
\end{figure}

In this paper, we theoretically demonstrate the potential for dramatically improving the fidelity of individual addressing and quantum gates by eliminating light shifts and inhomogeneity effects within the framework of three-photon excitation of Rydberg states.
Several of our previous works have highlighted distinctive features of three-photon schemes.
For instance, we have shown that both Doppler and recoil effects can be completely suppressed in the geometry of three laser beams with a zero sum of their wave vectors \cite{Ryabtsev2011}.
Figure~\ref{fig:Lukin}(a) shows the three-photon excitation scheme $5s_{1/2}\!\to\!5p_{3/2}\!\to\!6s_{1/2}\!\to\!np$ in $^{87}$Rb atoms, previously used in our experimental studies of Rydberg states \cite{Ryabtsev2016,Entin2013,Yakshina2018,Yakshina2020,Tretyakov2022,Beterov2023,Beterov2024}.
In our most recent work \cite{Beterov2024}, we observed three-photon Rabi oscillations in the Rydberg excitation of a single $^{87}$Rb atom in an optical dipole trap for the first time.
To observe these Rabi oscillations, we applied a strong resonant radiation on the second step, while the resonant radiations on the first and third steps were set to moderate levels.
Preliminary theoretical analysis had shown that a strong coupling in the second step induces ac Stark splitting of the intermediate levels $5P_{3/2}$ and $6S_{1/2}$, and introduces effective detunings on the intermediate transitions, ensuring coherence in the three-photon excitation.

To illustrate the key advantages of introducing the auxiliary intermediate $6s_{1/2}$ state, we examine the three-step ladder excitation of Fig.~\ref{fig:Lukin}(a) with strong laser coupling in the second step and compare its properties to the two-photon excitation scheme $5s_{1/2}\!\to\!6p_{3/2}\!\to\!ns_{1/2}$ of Fig.~\ref{fig:Lukin}(b), which is commonly used in quantum information processing experiments with Rydberg atoms \cite{Levine2018,Levine2019,Ebadi2021,Evered2023,Bluvstein2023}. 
The corresponding two-photon linkage diagram is depicted in Fig.~\ref{fig:Lukin}(c), where the laser Rabi frequencies $\Omega_{1,2}(r)$ are functions of the radial coordinate $r$ of coaxial laser beams [see Eq.~(\ref{eq:Gaussian}) below]. 
Far from single-photon resonance, $\delta _{1}\!\gg\!\Omega _{1,2}$, the spatially inhomogeneous two-photon Rabi frequency $\Omega (r)$ and the two-photon detuning $\Delta (r)$ [see Fig.~\ref{fig:Lukin}(c)] are given \cite{Grimm2000} by
\begin{align}
  \label{eq:2Rabi}
  \Omega \!=\! \Omega _{1}(r)\Omega _{2}(r)/(2\delta _{1}) ; \quad
  \Delta \!=\! [\Omega _{2}^{2}(r)-\Omega _{1}^{2}(r)]/(4\delta _{1}) \, . 
\end{align}

In further discussion of the three-photon scheme features, we denote states $5s_{1/2},5p_{3/2},6s_{1/2},np$ in Fig.~\ref{fig:Lukin}(a) as states$|1\rangle,|2\rangle, |3\rangle,|4\rangle$ respectively [see the corresponding linkage diagram in Fig.~\ref{fig:Lukin}(e)]. 
For each intermediate single-photon transition $j \!=\! 1,2,3$ we denote the respective detunings as $\delta _j$ and the Rabi frequencies as $\Omega _{j}\!=\! d_jE_j/\hbar$. 
Here $d_j$ are dipole moments of the single-photon transitions and $E_j$ are electric-field amplitudes of the linearly polarized light fields.
Due to the strong laser radiation ($\Omega _2\!\gg\! \Omega _{1,3}$) driving the second excitation step, the corresponding strong ac Stark splitting of the intermediate levels $5P_{3/2}$, $6S_{1/2}$ results in large effective detunings $\simeq \Omega _2/2$ [see Fig.~\ref{fig:Lukin}(d)], rendering these levels nearly unpopulated.
Therefore, as a reasonable approximation \cite{Shore2009}, instead of using the density matrix formalism to describe the temporal dynamics of a four-level system with radiatively decaying intermediate states, we can work in terms of the probability amplitudes $C_j(t)$ of the levels satisfying the reduced Schrödinger equation
\begin{equation}
  \begin{split}
    & i\dot{C}_1\!=\!\Omega _1C_2\text{e}^{i\delta _1t}/2 \,, 
    \\ 
    & i\dot{C}_2\!=\!\left(-iC_2/\tau _2 \!+\!\Omega _1C_1\text{e}^{-i\delta _1t} \!+\!\Omega _2C_3\text{e}^{i\delta _2t}\right)/2 \,, 
    \\ 
    & i\dot{C}_3\!=\!\left(-iC_3/\tau _3 \!+\!\Omega _2C_2\text{e}^{-i\delta _2t} \!+\!\Omega _3C_4\text{e}^{i\delta _3t}\right)/2 \,, 
    \\ 
    & i\dot{C}_4\!=\!\left(-iC_4/\tau _4 \!+\! \Omega _3C_3\text{e}^{-i\delta _3t} \right)/2 \,, 
  \end{split}
  \label{eq:SchrodingerI} 
\end{equation}
\noindent 
written in the rotating wave approximation. 
The radiative decays of both virtual intermediate and long-lived Rydberg states with radiative lifetimes $\tau _j$ are taken into account in Eq.~(\ref{eq:SchrodingerI}) by adding relaxation rate constants $1/\tau _j $ to the imaginary parts of the corresponding state energies \cite{Shore2009,CohenTannoudji1998,Grimm2000}.

The strong coupling of the intermediate states $|2\rangle,|3\rangle $, accompanied by their large splitting $\!\approx \!\Omega _2$, much larger than all other laser Rabi frequencies $\Omega_j$ and detunings $\delta _j$, allows an important simplification. 
Namely, for the amplitudes $C_{2,3}$ of the intermediate states one can apply the adiabatic elimination procedure \cite{stenholm2012foundations,Shore2009,Macr2023}, according to which the derivatives $\dot{C}_2,\dot{C}_3\!\approx\!0$, since their values are expected to be much smaller than the other terms. 
As a result, Eqs.~(\ref{eq:SchrodingerI}) are reduced to equations
\begin{equation}
  \begin{split}
    & i\dot{C}_1\!\approx \!\left(-i\overline{\gamma}_1 C_{1}\!-\! \Omega \!\cdot \! C_{4}\text{e}^{i\overline{\delta }t}\right )/2 \,, \\ 
    & i\dot{C}_4\!\approx \!\left(-i\overline{\gamma}_4 C_{4}\!-\! \Omega \!\cdot \!C_{1}\text{e}^{-i\overline{\delta }t}\right )/2 
  \end{split}
  \label{eq:SchrodingerR} 
\end{equation}
\noindent 
for an effective two-level system, involving only amplitudes $C_{1},C_{4}$. 
The three-photon detuning $\overline{\delta}$, reduced Rabi frequency $\Omega $ and relaxation constants $\overline{\gamma} _{1,4}$ have the following form:
\begin{equation}
  \begin{split}
    & \overline{\delta }\!=\! \delta _1\!+\!\delta _2\!+\!\delta _3;
      \quad
      \Omega \!=\!\Omega _{1}\Omega _{3}/\Omega _{2}; 
      \quad 
      \overline{\gamma} _1\!=\!\overline{\gamma}\!\cdot\!\Omega _{1}^2/\Omega _{2}^2;
    \\ 
    & \overline{\gamma} _4\!=\!\overline{\gamma}\!\cdot\!\Omega _{3}^2/\Omega _{2}^2\!+\!1/\tau _4; 
      \quad 
      \overline{\gamma}\!=\!(1/\tau _2\!+\!1/\tau _3)/2 . 
  \end{split}
  \label{eq:RabiRelaxEf} 
\end{equation}
\noindent 
Importantly, when $\tau _2 \!\neq \! \tau _3$ the conventional adiabatic elimination procedure ($\dot{C}_{2,3}\!\to\!0$) requires some corrections \cite{Macr2023}, which we have taken into account for the rate constants $\overline{\gamma} _{1,4}$ in Eq.~(\ref{eq:RabiRelaxEf}).

\begin{figure}
  \includegraphics[width=\linewidth]{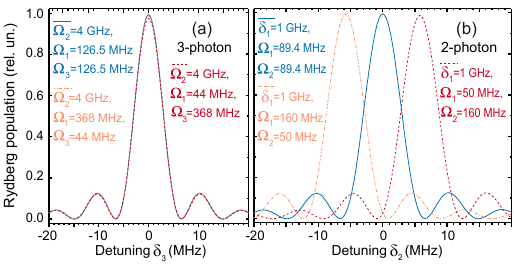}
  \caption{
    (a) Numerically obtained spectra of the three-photon excitation $5s_{1/2}\!\to\!5p_{3/2}\!\to\!6s_{1/2}\!\to\!70p_{3/2}$ [see Fig.~\ref{fig:Lukin}(a)] in $^{87}$Rb atoms for an interaction time of $0.125\,\mu$s at a high second-step laser Rabi frequency $\Omega _ 2/(2\pi)\!=\!4$ GHz and various ratios between the modest first-step $\Omega _ 1$ and third-step $\Omega _ 3$ frequencies, providing a fixed three-photon Rabi frequency $\Omega/(2\pi)\!=\!4$ MHz. 
    No light shift is observed for any combination of $\Omega _{1,3}$. 
    (b) The same for the two-photon excitation $5s_{1/2}\!\to\!6p_{3/2}\!\to\!70s_{1/2}$ [see Fig.~\ref{fig:Lukin}(b)] in $^{87}$Rb atoms for an interaction time of $0.125\,\mu$s at first-step detuning $\delta _1 /(2\pi)\!=\!1$ GHz and various ratios between the first-step $\Omega _ 1$ and second-step $\Omega _ 2$ laser frequencies, resulting in the same two-photon Rabi frequency $\Omega/(2\pi)\!=\!4$ MHz.
    A strong two-photon detuning $\Delta $ is observed when $\Omega_1\!\neq \!\Omega_2$ in agreement with Eq.~(\ref{eq:2Rabi}). 
    All spectra in both frames (a), (b) refer to the case of wide laser beams.
  }
  \label{fig:fig2}
\end{figure}

Since we are concerned with the case $\Omega _2 \!\gg \!\Omega _{1,3}$, $1/\tau _j$, $|\overline{\delta }|$, Eq.~(\ref{eq:RabiRelaxEf}) gives us an important set of inequalities:
\begin{align}
  \label{eq:2Relax}
  \overline{\gamma} _j \!= \!\frac{\overline{\gamma}}{\Omega _2}\frac{\Omega _j}{\Omega _{k\neq j}} \Omega \!\ll \!\Omega; 
  \quad 
  j,k\!= \!1,4 \, .
\end{align}
\noindent 
This significantly simplifies the formulaic representations of Eq.~(\ref{eq:SchrodingerR}) solutions along with the corresponding Rydberg states population $n_{4}\!=\!|C_4|^2$, which under the initial conditions $C_1(0)\!=\!1,C_{4}(0)\!=\!0$ acquires a compact form
\begin{equation}
  \begin{split}
      & n_{4}\!\approx\! \frac{\Omega ^2}{2\overline{\Omega}^2} \left [ 1\!-\!\cos\left (\overline{\Omega} \!\cdot\! t \right ) \right ]\text{e}^{-\Gamma t/2} ;
      \\ 
      & \overline{\Omega}\!=\!\sqrt{\Omega ^2\!+\!\overline{\delta }\, ^2};
        \quad 
        \Gamma\!=\!\frac{\Omega _{1}^2\!+\!\Omega _{3}^2}{\Omega _{2}^2}\overline{\gamma }\!+\!\frac{1}{\tau _{4}},
  \end{split}
  \label{eq:Population} 
\end{equation}
\noindent 
describing slowly fading coherent three-photon Rabi oscillations at frequency $\overline{\Omega} $. 
The first term in the decay constant $\Gamma $ of Eq.~(\ref{eq:Population}) accounts for decoherence due to a small partial population of the short-lived intermediate states $|2\rangle $, $|3\rangle $, while the second term is due to the long lifetime of the Rydberg state $|4\rangle $. 
In the following discussion we mainly focus on the case of resonant lasers with $\delta _{1,2,3}\!=\!0$, i.e when the three-photon detuning $\overline{\delta }\!=\!0$ and $\overline{\Omega}\!=\!\Omega$. 
The one exception is the spectral profiles in Fig.~\ref{fig:fig2}(a), obtained by varying $\delta _3$ around zero.

Noteworthy, the three-photon Rabi frequency~(\ref{eq:RabiRelaxEf}) takes a form similar to the two-photon case~(\ref{eq:2Rabi}), with the Rabi frequency $\Omega_2/2$ taking the role of the single-photon detuning $\delta_1$ in Fig.~\ref{fig:Lukin}(c).
The large $\Omega _2$ in Fig.~\ref{fig:Lukin}(e) forms two adiabatic states $|\pm \rangle \!= \!(|2 \rangle \!\pm \!|3 \rangle)/\sqrt{2}$, thus virtually converting the three-photon excitation into a double two-photon scheme depicted in Fig.~\ref{fig:Lukin}(d) by inducing light shifts $\pm \Omega _2/2$ with two fundamentaly new properties:

\textbf{First}, if we arrange the spatial distributions of $(\Omega _{1}\Omega _{3})$ and $\Omega _{2}$ to be identical, then $\Omega$ becomes independent of the atom position, even with tightly focused laser beams. 
This enables a significantly improved individual addressing of Rydberg excitations for adjacent atoms in trap arrays, compared to two-photon excitation schemes. 
In addition, tight focusing strongly reduces the total laser power required to excite Rydberg atoms in large-scale quantum registers of neutral atoms.

\textbf{Second}, the essential trait of the reduced Eq.~(\ref{eq:SchrodingerR}) is the total elimination of ac Stark shifts $\Delta _{j=1,4}$ for both the ground and Rydberg states. 
Because of the mirror-image arrangement of the virtual levels $|\pm \rangle $ in the linkage diagram, associated with Eq.~(\ref{eq:SchrodingerI}) and shown in Fig.~\ref{fig:Lukin}(d), they cause two partial identical but opposite optical shifts $\Delta _{j\pm}$ of each \textit{j}-state, making the overall shift $\Delta _j$ zero at arbitrary laser intensities. 
As a result, the three-photon detuning $\Delta \!=\!\Delta _4\!-\!\Delta _1$ turns out to be insensitive to any spatial inhomogeneities of the laser fields. 
Therefore, even in tightly focused laser beams with radically different $\Omega_1$ and $\Omega_3$ values, the spatially inhomogeneous light shifts will be absent. 
The latter can significantly increase the coherence and fidelity of Rydberg quantum gates.

The above simple analysis highlights the main advantages of using three-photon laser excitation of Rydberg states for both individual addressing and light-shift suppression. 
Next we demonstrate how these advantages are manifested in accurate numerical calculations of light shifts and Rabi oscillations.

A full rigorous theoretical analysis must deal with the density matrix formalism and the optical Bloch equations \cite{Shore2009,stenholm2012foundations}, traditionally used to describe atom-light interaction processes, and taking into account the real hyperfine and Zeeman structures of all states involved in the interaction. 
In our previous works \cite{Kirova2017,PhysRevA.109.063116} we developed an efficient algorithm based on the split operator technique \cite{Kazansky,efimov2014}, which provides robust symplectic \cite{Hairer2006,Hairer2003} numerical simulations of laser excitation of alkali metal atoms. 
Details of the algorithm specific parameters as applied to Rb and Na atoms can be found in \cite{PhysRevA.77.042511,PhysRevA.109.063116}.

Figure~\ref{fig:fig2} presents results of our numerical simulations of the spectra profiles for the cases of (a) three-photon and (b) two-photon Rydberg states excitation in $^{87}$Rb atoms for a fixed interaction time of $0.125\,\mu$s. 
The radiative lifetimes of the Rydberg states at T=300 K are $\tau(70p _{3/2})=190\,\mu$s and $\tau(70s _{1/2})=152\,\mu$s, respectively. 
The lifetimes of intermediate states are schematically shown in Fig.~\ref{fig:Lukin}.
The strong coupling of intermediate states in case (a) and the large single-photon detuning $\delta_1$ in case (b) make all the intermediate states virtual.
Our numerical simulations clearly demonstrate in Fig.~\ref{fig:fig2}(a) the absence of ac Stark shifts for arbitrary values of the laser Rabi frequencies $\Omega_{1,3}$. 
In contrast, Fig.~\ref{fig:fig2}(b) demonstrates a strong dependence of the two-photon resonance position on the imbalance parameter $\Omega_{2}/\Omega_{1}$ between the Rabi frequencies. 
Therefore, spatial inhomogeneity or intensity fluctuations of the laser radiations would strongly affect two-photon excitation, while three-photon excitation is robust against these factors.

\begin{figure}
  \includegraphics[width=\linewidth]{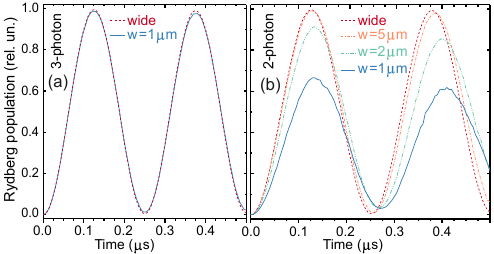}
  \caption{
    (a) Rabi oscillations simulations for three-photon excitation $5s_{1/2}\!\to\!5p_{3/2}\!\to\!6s_{1/2}\!\to\!70p_{3/2}$ in $^{87}$Rb atoms with strong second-step laser coupling ${\Omega_2/(2\pi)\!=\!4}$~GHz and moderate first and third-step Rabi frequencies $\Omega_1/(2\pi)\!=\!\Omega_3/(2\pi)\!=\!126.5$~MHz. 
    The dashed red curve corresponds to spatially uniform (wide) laser beams, while the solid blue curve describes the spatially averaged oscillations $\langle n_4 \rangle $ at $a\!=\!1\,\mu$m and $w\!=\!1\,\mu$m for the radii of the atomic $a$ (\ref{eq:Distribution}) and the laser $w$ (\ref{eq:RabiStable}) spots, respectively.
    Spatial variation of Rabi frequencies $\Omega_j(r)$ does not noticeably affect contrast of the curve. 
    (b) The same for the two-photon excitation $5s_{1/2}\!\to\!6p_{3/2}\!\to\!70s_{1/2}$ in $^{87}$Rb atoms for single-photon detuning $\delta _1 /(2\pi)\!=\!1$~GHz and first- and second-step Rabi frequencies $\Omega _1/(2\pi)\!=\!160$~MHz, $\Omega _2/(2\pi)\!=\!50$~MHz, respectively, as used in \cite{Levine2019}. 
    Laser spot radius varies from a very large value (wide beams) to $w\!=\!5, 2, 1\,\mu$m. 
    The spatial variations of Rabi frequencies $\Omega _{1,2}(r)$ strongly reduce the contrast in tightly focused laser beams.
  }
  \label{fig:fig3}
\end{figure}

\begin{figure}
  \includegraphics[width=\linewidth]{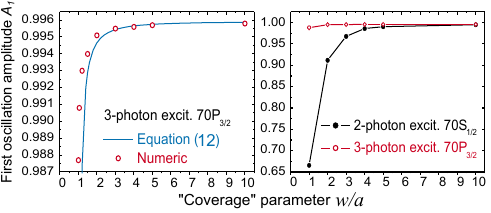}
  \caption{ 
    (a) Numerically (circles) and analytically (blue curve) obtained spatially averaged amplitude $A_1$ of the first Rabi oscillation at three-photon excitation $5s_{1/2}\!\to\!5p_{3/2}\!\to\!6s_{1/2}\!\to\!70p_{3/2}$ in $^{87}$Rb atoms, for the same parameters as in Fig.~3(a).
    The circles and the curve demonstrate the dependence of amplitude $A_1$ on the ``coverage'' parameter $w/a$. 
    Fairly good agreement is observed for $w/a \!\geq\! 2$.
    (b) Comparison of numerically calculated spatially averaged amplitude $A_1$ for three-photon excitation corresponding to frame (a) and two-photon excitation $5s_{1/2}\!\to\!6p_{3/2}\!\to\!70s_{1/2}$ in $^{87}$Rb atoms, for the same parameters as in Fig.~ 3(b). 
    Under three-photon excitation, contrast of the first peak remains virtually unchanged even in fairly narrow laser beams with a coverage parameter of $w/a\!=\!2$. 
    In contrast, the two-photon oscillations exhibit a significant drop of $A_1$ amplitude in moderately narrow beams with $w/a\!\leq\!5$.
  }
  \label{fig:4}
\end{figure}

The ability to achieve a spatially homogeneous Rabi frequency $\Omega $~(\ref{eq:RabiRelaxEf}) with a special laser beam arrangement is the most distinctive feature of the three-photon scheme.
This configuration enables highly coherent excitation of Rydberg states even in very tightly focused beams, as illustrated in Figs.~\ref{fig:fig3},~\ref{fig:4}.
In cylindrical coordinates $r, \varphi$, the coaxial laser beams have Gaussian Rabi frequency profiles with respect to the transverse radial coordinate 
\begin{align}
  \label{eq:Gaussian}
  \Omega _{j}(r) \!=\! \Omega _{j}\text{e}^{-r^2/w_j^2 }; \quad \Omega _{j}\!\equiv \!\Omega _{j}(r\!=\!0).
\end{align}
\noindent 
The parameters $w_ j$ ($j\!=\!1-3$) represent the $1/\textnormal{e}^{2}$ laser focal spot radii, usually measured in experiments. 
The following relationship
\begin{align}
  \label{eq:RabiStable}
  w_1\!=\!w_3\!=\! \sqrt{2}w_2\!\equiv\! w; \, \Rightarrow \, 
  \Omega (r)\!=\!\frac{\Omega _{1}(r\!=\!0)\Omega _{3}(r\!=\!0)}{\Omega _{2}(r\!=\!0)}
\end{align}
\noindent 
between laser spots guarantees the constancy of the three-photon Rabi oscillation frequency for any spatial position of the atom. 
This is a very important finding, since the atom in an optical dipole trap is not localized point-wise. 
Instead, the atom density has some azimuthally invariant radial distribution $\rho(r)$, reasonably well approximated by a normalized Gaussian function \cite{Grimm2000}
\begin{align}
  \label{eq:Distribution}
  \rho(r)\!=\!\frac{2}{\pi a^2}\text{e}^{-2r^2/a^2}. 
\end{align} 
\noindent 
The parameter $a$ is the atom spot radius for the probability level $1/\textnormal{e}^2$. 
It is determined by the waist radius $w_0$ of the optical dipole trap, atom temperature $T$, and trap depth $U_0$, and can be estimated as $a \!\sim \! w_0 \sqrt{k_B T / U_0}$ \cite{Grimm2000}, where $k_B$ is the Boltzmann constant. 
For the typical trap depth $ \! \sim \! 1$~mK and atom temperature $ \! \sim \! 10$ $\mu$K, one has $a \!\sim \! w_0/10$.

The Rabi oscillations observed in an experiment represent the average population $\langle n_4 \rangle $ (\ref{eq:Population}) over spatial coordinates $r, \varphi$, with the weight function $\rho(r)$ (\ref{eq:Distribution}). 
Figure~\ref{fig:fig3} presents numerically calculated Rabi oscillations of (a)~three-photon and (b)~two-photon excitation in $^{87}$Rb atoms. 
The data in frame~(a) are calculated for a large laser Rabi frequency in the second step, and equal moderate Rabi frequencies applied to the first and third excitation steps. 
The two-photon oscillation curves in frame~(b) were calculated using the single-photon detuning and the first- and second-step Rabi frequencies values from \cite{Levine2019}.
The dashed red curves in both frames of Fig.~\ref{fig:fig3} correspond to spatially uniform (wide) laser beams. 
Both curves have the same three- and two-photon Rabi oscillation frequencies of 4 MHz and identical oscillation contrasts. 
The other curves in Fig.~\ref{fig:fig3} show how the spatially averaged Rabi oscillations depend on different relationships between the uncertainty of the atomic position and the laser spot size. 
The curves indicate a completely different character of the destructive influence of spatial inhomogeneities of laser fields on the oscillation contrast for two- and three-step schemes. 
In a more detailed discussion below, we will obtain a formulaic description for the effects of spatial variation in laser intensities.

For our three-photon excitation, the radial dependence of the Rydberg state population (\ref{eq:Population}) arises only due to variation in the effective relaxation constant $\Gamma $ included in the amplitude part of Eq.~(\ref{eq:Population}):
\begin{align}
  \label{eq:Relax}
  \Gamma(r)\!=\!\frac{\Omega _{1}^2(0)\!+\!\Omega _{3}^2(0)}{2\Omega _{2}^2(0)}\left(\frac{1}{\tau _2}\!+\frac{1}{\tau _3}\right)\text{e}^{2r^2/w^2}\!+\!\frac{1}{\tau _{4}} \, ,
\end{align}
\noindent 
which is a consequence of the relation~(\ref{eq:RabiStable}) for laser focal spots. 
Importantly, the population averaging procedure affects only the relaxation factor, leaving the oscillatory part in Eq.~(\ref{eq:Population}) unchanged.

The main practically important parameter of Rabi oscillations, which directly defines the maximum achievable fidelity of Rydberg quantum gates, is the amplitude $A_1$ of the first oscillation peak in Eq.~(\ref{eq:Population}). 
The latter occurs at time $t\!=\!\pi/\Omega$. 
For a given spatial point with coordinate $r$ and for $\Omega \!\gg\!\Gamma (r)$, the peak height is expressed as
\begin{align}
  \label{eq:Peak}
  n_4(r)_{max} \!=\! \text{e}^{-\Gamma(r) t/2} \! \approx\!1 \!-\!\pi \Gamma(r)/(2\Omega) \, . 
\end{align}
\noindent 
Averaging of Eq.~(\ref{eq:Peak}) over cylindrical spacial coordinates $r,\varphi$ with the weight function~(\ref{eq:Distribution}) yields 
\begin{align} 
  \label{eq:AveragePeak}
  A_1\!\equiv \!\left< n_4(t)\right> _{max}\approx\!1 \!-\!\frac{\pi }{2\Omega}
  \left[
    \frac{\Gamma(0)\xi ^2}{\xi ^2\!-\!1} \!-\!\frac{\tau _4^{-1}}{\xi ^2\!-\!1}
  \right],
\end{align}
\noindent 
where the ``coverage'' parameter $\xi \!=\!w/a$ shows how much the Rydberg laser spot exceeds (``covers'') the atom spot.

Figure~\ref{fig:4}(a) shows the numerically identified (circles) and analytically calculated using Eq.~(\ref{eq:AveragePeak}) (blue curve) dependences of the spatially averaged amplitude $A_1$ on the ratio $w/a$ for three-photon excitation in $^{87}$Rb atoms, with the same parameters as in Fig.~\ref{fig:fig3}(a). 
A fairly good agreement between the analytical model and numerical simulation is observed for $w/a \!\geq\! 2$. 
We also see a slow decline in $A_1$ in the interval $1\! \leq \! w/a\! \leqslant \!2 $, where the amplitude remains close to its maximum possible value of $0.9957$. 
Moreover, even for $w/a \!=\! 1$ the value of $A_1$ drops to only slightly lower value of $0.9877$. 
This change in $A_1$ is almost imperceptible in the Rabi oscillations shown in Fig.~\ref{fig:fig3}(a). 
Our simulations confirm the validity of the above proposal for high precision addressing with three-photon Rydberg excitation, and show that it should be experimentally feasible.


The obtained results open up prospects for an experimental solution to the problem of local addressing in atomic arrays. 
A typical experiment with single atoms in quantum registers assumes the values of atomic and laser spots to be $a \! \sim \! 1\,\mu$m and $w \! \sim \!2 \!- \!3\,\mu$m, respectively, and the distance between neighboring atoms of $ \! \sim \! 5\,\mu$m. 
At the Rabi frequencies $\Omega _j$ used in Fig.~\ref{fig:4}(a), the calculated amplitude $A_1$ of the first oscillation is $0.9951$ at $w/a\! = \!2$. 
The significant exponential drop of $\Omega _j$ in Eq.~(\ref{eq:Gaussian}) strongly suppresses the Rydberg excitation of the neighboring atoms: the calculations yield a negligible value of $ \! \sim \!10^{ -5}\! -\!10^ {-6}$ for their Rydberg state populations.

The above observation is not the case for two-photon excitation. 
Figure~\ref{fig:4}(b) shows the comparison of numerically calculated spatially-averaged amplitudes $A_1$ of the first Rabi oscillation at three-photon excitation and two-photon excitation in $^{87}$Rb atoms for the same parameters as in Fig.~\ref{fig:fig3}(b). 
While the three-photon amplitude $A_1$ in Fig.~\ref{fig:4}(b) is almost independent of the change in $w/a$, the two-photon amplitude decreases significantly starting from $w/a\! =\!5$. 
At $w/a\! =\!1$ it drops to $0.662$ compared to $0.9949$ for $w/a\! =\!10$ and $0.9956$ for wide laser beams. 
At $w/a\! =\!2$ the two-photon amplitude $A_1$ takes the value $0.912$, which is no longer suitable for quantum gates.
Therefore, precise individual addressing cannot be implemented with two-photon Rydberg excitation in atom arrays with spacing of $\! \sim \! 5$ $\mu$m. 
To avoid crosstalk problems, the distance should be increased to at least $\! \sim \! 20$ $\mu$m, while using wide laser beams with $w/a\! \sim \!10$. 
This prevents implementation of large-scale quantum registers with neutral atoms, where the smallest possible distance between adjacent atoms is necessary to ensure strong interaction of Rydberg states and compact packing of atomic arrays.

In conclusion, we have shown theoretically that high-fidelity individual addressing of single atoms and the mitigation of crosstalk in optical dipole trap arrays can be achieved using three-photon laser excitation of Rydberg states, provided that strong laser radiation in the second step induces a dominant ac Stark splitting of the two intermediate states. 
Moreover, this excitation method suppresses the light shift of the three-photon resonance, even in cases of arbitrary imbalance between the Rabi frequencies in the first and third steps.
Thus, three-photon Rydberg excitation offers a substantial advantage over the commonly used two-photon excitation, in which individual addressing and light-shift suppression are challenging to implement experimentally.

Authors N.N.B., I.I.B., and I.I.R. acknowledge the support of the grant No.~23-12-00067 (https://rscf.ru/project/23-12-00067/) by the Russian Science Foundation.
Author A.C. acknowledges the support of Latvian Council of Science project No.~{lzp-2023/1-0199}.


\bibliography{PRL-main}

\begin{thebibliography}{45}%
\makeatletter
\providecommand \@ifxundefined [1]{%
 \@ifx{#1\undefined}
}%
\providecommand \@ifnum [1]{%
 \ifnum #1\expandafter \@firstoftwo
 \else \expandafter \@secondoftwo
 \fi
}%
\providecommand \@ifx [1]{%
 \ifx #1\expandafter \@firstoftwo
 \else \expandafter \@secondoftwo
 \fi
}%
\providecommand \natexlab [1]{#1}%
\providecommand \enquote  [1]{``#1''}%
\providecommand \bibnamefont  [1]{#1}%
\providecommand \bibfnamefont [1]{#1}%
\providecommand \citenamefont [1]{#1}%
\providecommand \href@noop [0]{\@secondoftwo}%
\providecommand \href [0]{\begingroup \@sanitize@url \@href}%
\providecommand \@href[1]{\@@startlink{#1}\@@href}%
\providecommand \@@href[1]{\endgroup#1\@@endlink}%
\providecommand \@sanitize@url [0]{\catcode `\\12\catcode `\$12\catcode
  `\&12\catcode `\#12\catcode `\^12\catcode `\_12\catcode `\%12\relax}%
\providecommand \@@startlink[1]{}%
\providecommand \@@endlink[0]{}%
\providecommand \url  [0]{\begingroup\@sanitize@url \@url }%
\providecommand \@url [1]{\endgroup\@href {#1}{\urlprefix }}%
\providecommand \urlprefix  [0]{URL }%
\providecommand \Eprint [0]{\href }%
\providecommand \doibase [0]{https://doi.org/}%
\providecommand \selectlanguage [0]{\@gobble}%
\providecommand \bibinfo  [0]{\@secondoftwo}%
\providecommand \bibfield  [0]{\@secondoftwo}%
\providecommand \translation [1]{[#1]}%
\providecommand \BibitemOpen [0]{}%
\providecommand \bibitemStop [0]{}%
\providecommand \bibitemNoStop [0]{.\EOS\space}%
\providecommand \EOS [0]{\spacefactor3000\relax}%
\providecommand \BibitemShut  [1]{\csname bibitem#1\endcsname}%
\let\auto@bib@innerbib\@empty
\bibitem [{\citenamefont {Saffman}\ \emph {et~al.}(2010)\citenamefont
  {Saffman}, \citenamefont {Walker},\ and\ \citenamefont
  {M\o{}lmer}}]{Saffman2010}%
  \BibitemOpen
  \bibfield  {author} {\bibinfo {author} {\bibfnamefont {M.}~\bibnamefont
  {Saffman}}, \bibinfo {author} {\bibfnamefont {T.~G.}\ \bibnamefont
  {Walker}},\ and\ \bibinfo {author} {\bibfnamefont {K.}~\bibnamefont
  {M\o{}lmer}},\ }\bibfield  {title} {\bibinfo {title} {Quantum information
  with rydberg atoms},\ }\href {https://doi.org/10.1103/RevModPhys.82.2313}
  {\bibfield  {journal} {\bibinfo  {journal} {Rev. Mod. Phys.}\ }\textbf
  {\bibinfo {volume} {82}},\ \bibinfo {pages} {2313} (\bibinfo {year}
  {2010})}\BibitemShut {NoStop}%
\bibitem [{\citenamefont {Saffman}(2016)}]{Saffman2016}%
  \BibitemOpen
  \bibfield  {author} {\bibinfo {author} {\bibfnamefont {M.}~\bibnamefont
  {Saffman}},\ }\bibfield  {title} {\bibinfo {title} {Quantum computing with
  atomic qubits and rydberg interactions: progress and challenges},\ }\href
  {https://doi.org/10.1088/0953-4075/49/20/202001} {\bibfield  {journal}
  {\bibinfo  {journal} {Journal of Physics B: Atomic, Molecular and Optical
  Physics}\ }\textbf {\bibinfo {volume} {49}},\ \bibinfo {pages} {202001}
  (\bibinfo {year} {2016})}\BibitemShut {NoStop}%
\bibitem [{\citenamefont {Ryabtsev}\ \emph {et~al.}(2016)\citenamefont
  {Ryabtsev}, \citenamefont {Beterov}, \citenamefont {Tret’yakov},
  \citenamefont {Èntin},\ and\ \citenamefont {Yakshina}}]{Ryabtsev2016}%
  \BibitemOpen
  \bibfield  {author} {\bibinfo {author} {\bibfnamefont {I.~I.}\ \bibnamefont
  {Ryabtsev}}, \bibinfo {author} {\bibfnamefont {I.~I.}\ \bibnamefont
  {Beterov}}, \bibinfo {author} {\bibfnamefont {D.~B.}\ \bibnamefont
  {Tret’yakov}}, \bibinfo {author} {\bibfnamefont {V.~M.}\ \bibnamefont
  {Èntin}},\ and\ \bibinfo {author} {\bibfnamefont {E.~A.}\ \bibnamefont
  {Yakshina}},\ }\bibfield  {title} {\bibinfo {title} {Spectroscopy of cold
  rubidium rydberg atoms for applications in quantum information},\ }\href
  {https://doi.org/10.3367/ufne.0186.201602k.0206} {\bibfield  {journal}
  {\bibinfo  {journal} {Physics-Uspekhi}\ }\textbf {\bibinfo {volume} {59}},\
  \bibinfo {pages} {196–208} (\bibinfo {year} {2016})}\BibitemShut {NoStop}%
\bibitem [{\citenamefont {Henriet}\ \emph {et~al.}(2020)\citenamefont
  {Henriet}, \citenamefont {Beguin}, \citenamefont {Signoles}, \citenamefont
  {Lahaye}, \citenamefont {Browaeys}, \citenamefont {Reymond},\ and\
  \citenamefont {Jurczak}}]{Henriet2020}%
  \BibitemOpen
  \bibfield  {author} {\bibinfo {author} {\bibfnamefont {L.}~\bibnamefont
  {Henriet}}, \bibinfo {author} {\bibfnamefont {L.}~\bibnamefont {Beguin}},
  \bibinfo {author} {\bibfnamefont {A.}~\bibnamefont {Signoles}}, \bibinfo
  {author} {\bibfnamefont {T.}~\bibnamefont {Lahaye}}, \bibinfo {author}
  {\bibfnamefont {A.}~\bibnamefont {Browaeys}}, \bibinfo {author}
  {\bibfnamefont {G.-O.}\ \bibnamefont {Reymond}},\ and\ \bibinfo {author}
  {\bibfnamefont {C.}~\bibnamefont {Jurczak}},\ }\bibfield  {title} {\bibinfo
  {title} {Quantum computing with neutral atoms},\ }\href
  {https://doi.org/10.22331/q-2020-09-21-327} {\bibfield  {journal} {\bibinfo
  {journal} {{Quantum}}\ }\textbf {\bibinfo {volume} {4}},\ \bibinfo {pages}
  {327} (\bibinfo {year} {2020})}\BibitemShut {NoStop}%
\bibitem [{\citenamefont {Shi}(2022)}]{Shi2022}%
  \BibitemOpen
  \bibfield  {author} {\bibinfo {author} {\bibfnamefont {X.-F.}\ \bibnamefont
  {Shi}},\ }\bibfield  {title} {\bibinfo {title} {Quantum logic and
  entanglement by neutral rydberg atoms: methods and fidelity},\ }\href
  {https://doi.org/10.1088/2058-9565/ac18b8} {\bibfield  {journal} {\bibinfo
  {journal} {Quantum Science and Technology}\ }\textbf {\bibinfo {volume}
  {7}},\ \bibinfo {pages} {023002} (\bibinfo {year} {2022})}\BibitemShut
  {NoStop}%
\bibitem [{\citenamefont {Barredo}\ \emph {et~al.}(2016)\citenamefont
  {Barredo}, \citenamefont {de~Léséleuc}, \citenamefont {Lienhard},
  \citenamefont {Lahaye},\ and\ \citenamefont {Browaeys}}]{Barredo2016}%
  \BibitemOpen
  \bibfield  {author} {\bibinfo {author} {\bibfnamefont {D.}~\bibnamefont
  {Barredo}}, \bibinfo {author} {\bibfnamefont {S.}~\bibnamefont
  {de~Léséleuc}}, \bibinfo {author} {\bibfnamefont {V.}~\bibnamefont
  {Lienhard}}, \bibinfo {author} {\bibfnamefont {T.}~\bibnamefont {Lahaye}},\
  and\ \bibinfo {author} {\bibfnamefont {A.}~\bibnamefont {Browaeys}},\
  }\bibfield  {title} {\bibinfo {title} {An atom-by-atom assembler of
  defect-free arbitrary two-dimensional atomic arrays},\ }\href
  {https://doi.org/10.1126/science.aah3778} {\bibfield  {journal} {\bibinfo
  {journal} {Science}\ }\textbf {\bibinfo {volume} {354}},\ \bibinfo {pages}
  {1021} (\bibinfo {year} {2016})}\BibitemShut {NoStop}%
\bibitem [{\citenamefont {Pause}\ \emph {et~al.}(2024)\citenamefont {Pause},
  \citenamefont {Sturm}, \citenamefont {Mittenb\"{u}hler}, \citenamefont
  {Amann}, \citenamefont {Preuschoff}, \citenamefont {Sch\"{a}ffner},
  \citenamefont {Schlosser},\ and\ \citenamefont {Birkl}}]{Pause2024}%
  \BibitemOpen
  \bibfield  {author} {\bibinfo {author} {\bibfnamefont {L.}~\bibnamefont
  {Pause}}, \bibinfo {author} {\bibfnamefont {L.}~\bibnamefont {Sturm}},
  \bibinfo {author} {\bibfnamefont {M.}~\bibnamefont {Mittenb\"{u}hler}},
  \bibinfo {author} {\bibfnamefont {S.}~\bibnamefont {Amann}}, \bibinfo
  {author} {\bibfnamefont {T.}~\bibnamefont {Preuschoff}}, \bibinfo {author}
  {\bibfnamefont {D.}~\bibnamefont {Sch\"{a}ffner}}, \bibinfo {author}
  {\bibfnamefont {M.}~\bibnamefont {Schlosser}},\ and\ \bibinfo {author}
  {\bibfnamefont {G.}~\bibnamefont {Birkl}},\ }\bibfield  {title} {\bibinfo
  {title} {Supercharged two-dimensional tweezer array with more than 1000
  atomic qubits},\ }\href {https://doi.org/10.1364/OPTICA.513551} {\bibfield
  {journal} {\bibinfo  {journal} {Optica}\ }\textbf {\bibinfo {volume} {11}},\
  \bibinfo {pages} {222} (\bibinfo {year} {2024})}\BibitemShut {NoStop}%
\bibitem [{\citenamefont {Pichard}\ \emph {et~al.}(2024)\citenamefont
  {Pichard}, \citenamefont {Lim}, \citenamefont {Bloch}, \citenamefont
  {Vaneecloo}, \citenamefont {Bourachot}, \citenamefont {Both}, \citenamefont
  {M\'eriaux}, \citenamefont {Dutartre}, \citenamefont {Hostein}, \citenamefont
  {Paris}, \citenamefont {Ximenez}, \citenamefont {Signoles}, \citenamefont
  {Browaeys}, \citenamefont {Lahaye},\ and\ \citenamefont
  {Dreon}}]{Pichard2024}%
  \BibitemOpen
  \bibfield  {author} {\bibinfo {author} {\bibfnamefont {G.}~\bibnamefont
  {Pichard}}, \bibinfo {author} {\bibfnamefont {D.}~\bibnamefont {Lim}},
  \bibinfo {author} {\bibfnamefont {E.}~\bibnamefont {Bloch}}, \bibinfo
  {author} {\bibfnamefont {J.}~\bibnamefont {Vaneecloo}}, \bibinfo {author}
  {\bibfnamefont {L.}~\bibnamefont {Bourachot}}, \bibinfo {author}
  {\bibfnamefont {G.-J.}\ \bibnamefont {Both}}, \bibinfo {author}
  {\bibfnamefont {G.}~\bibnamefont {M\'eriaux}}, \bibinfo {author}
  {\bibfnamefont {S.}~\bibnamefont {Dutartre}}, \bibinfo {author}
  {\bibfnamefont {R.}~\bibnamefont {Hostein}}, \bibinfo {author} {\bibfnamefont
  {J.}~\bibnamefont {Paris}}, \bibinfo {author} {\bibfnamefont
  {B.}~\bibnamefont {Ximenez}}, \bibinfo {author} {\bibfnamefont
  {A.}~\bibnamefont {Signoles}}, \bibinfo {author} {\bibfnamefont
  {A.}~\bibnamefont {Browaeys}}, \bibinfo {author} {\bibfnamefont
  {T.}~\bibnamefont {Lahaye}},\ and\ \bibinfo {author} {\bibfnamefont
  {D.}~\bibnamefont {Dreon}},\ }\bibfield  {title} {\bibinfo {title}
  {Rearrangement of individual atoms in a 2000-site optical-tweezer array at
  cryogenic temperatures},\ }\href
  {https://doi.org/10.1103/PhysRevApplied.22.024073} {\bibfield  {journal}
  {\bibinfo  {journal} {Phys. Rev. Appl.}\ }\textbf {\bibinfo {volume} {22}},\
  \bibinfo {pages} {024073} (\bibinfo {year} {2024})}\BibitemShut {NoStop}%
\bibitem [{\citenamefont {Manetsch}\ \emph {et~al.}(2024)\citenamefont
  {Manetsch}, \citenamefont {Nomura}, \citenamefont {Bataille}, \citenamefont
  {Leung}, \citenamefont {Lv},\ and\ \citenamefont {Endres}}]{manetsch2024}%
  \BibitemOpen
  \bibfield  {author} {\bibinfo {author} {\bibfnamefont {H.~J.}\ \bibnamefont
  {Manetsch}}, \bibinfo {author} {\bibfnamefont {G.}~\bibnamefont {Nomura}},
  \bibinfo {author} {\bibfnamefont {E.}~\bibnamefont {Bataille}}, \bibinfo
  {author} {\bibfnamefont {K.~H.}\ \bibnamefont {Leung}}, \bibinfo {author}
  {\bibfnamefont {X.}~\bibnamefont {Lv}},\ and\ \bibinfo {author}
  {\bibfnamefont {M.}~\bibnamefont {Endres}},\ }\href
  {https://arxiv.org/abs/2403.12021} {\bibinfo {title} {A tweezer array with
  6100 highly coherent atomic qubits}} (\bibinfo {year} {2024}),\ \Eprint
  {https://arxiv.org/abs/2403.12021} {arXiv:2403.12021 [quant-ph]} \BibitemShut
  {NoStop}%
\bibitem [{\citenamefont {Sheng}\ \emph {et~al.}(2018)\citenamefont {Sheng},
  \citenamefont {He}, \citenamefont {Xu}, \citenamefont {Guo}, \citenamefont
  {Wang}, \citenamefont {Xiong}, \citenamefont {Liu}, \citenamefont {Wang},\
  and\ \citenamefont {Zhan}}]{Sheng2018}%
  \BibitemOpen
  \bibfield  {author} {\bibinfo {author} {\bibfnamefont {C.}~\bibnamefont
  {Sheng}}, \bibinfo {author} {\bibfnamefont {X.}~\bibnamefont {He}}, \bibinfo
  {author} {\bibfnamefont {P.}~\bibnamefont {Xu}}, \bibinfo {author}
  {\bibfnamefont {R.}~\bibnamefont {Guo}}, \bibinfo {author} {\bibfnamefont
  {K.}~\bibnamefont {Wang}}, \bibinfo {author} {\bibfnamefont {Z.}~\bibnamefont
  {Xiong}}, \bibinfo {author} {\bibfnamefont {M.}~\bibnamefont {Liu}}, \bibinfo
  {author} {\bibfnamefont {J.}~\bibnamefont {Wang}},\ and\ \bibinfo {author}
  {\bibfnamefont {M.}~\bibnamefont {Zhan}},\ }\bibfield  {title} {\bibinfo
  {title} {High-fidelity single-qubit gates on neutral atoms in a
  two-dimensional magic-intensity optical dipole trap array},\ }\href
  {https://doi.org/10.1103/PhysRevLett.121.240501} {\bibfield  {journal}
  {\bibinfo  {journal} {Phys. Rev. Lett.}\ }\textbf {\bibinfo {volume} {121}},\
  \bibinfo {pages} {240501} (\bibinfo {year} {2018})}\BibitemShut {NoStop}%
\bibitem [{\citenamefont {Nikolov}\ \emph {et~al.}(2023)\citenamefont
  {Nikolov}, \citenamefont {Diamond-Hitchcock}, \citenamefont {Bass},
  \citenamefont {Spong},\ and\ \citenamefont {Pritchard}}]{Nikolov2023}%
  \BibitemOpen
  \bibfield  {author} {\bibinfo {author} {\bibfnamefont {B.}~\bibnamefont
  {Nikolov}}, \bibinfo {author} {\bibfnamefont {E.}~\bibnamefont
  {Diamond-Hitchcock}}, \bibinfo {author} {\bibfnamefont {J.}~\bibnamefont
  {Bass}}, \bibinfo {author} {\bibfnamefont {N.~L.~R.}\ \bibnamefont {Spong}},\
  and\ \bibinfo {author} {\bibfnamefont {J.~D.}\ \bibnamefont {Pritchard}},\
  }\bibfield  {title} {\bibinfo {title} {Randomized benchmarking using
  nondestructive readout in a two-dimensional atom array},\ }\href
  {https://doi.org/10.1103/PhysRevLett.131.030602} {\bibfield  {journal}
  {\bibinfo  {journal} {Phys. Rev. Lett.}\ }\textbf {\bibinfo {volume} {131}},\
  \bibinfo {pages} {030602} (\bibinfo {year} {2023})}\BibitemShut {NoStop}%
\bibitem [{\citenamefont {Ma}\ \emph {et~al.}(2023)\citenamefont {Ma},
  \citenamefont {Liu}, \citenamefont {Peng}, \citenamefont {Zhang},
  \citenamefont {Jandura}, \citenamefont {Claes}, \citenamefont {Burgers},
  \citenamefont {Pupillo}, \citenamefont {Puri},\ and\ \citenamefont
  {Thompson}}]{Ma2023}%
  \BibitemOpen
  \bibfield  {author} {\bibinfo {author} {\bibfnamefont {S.}~\bibnamefont
  {Ma}}, \bibinfo {author} {\bibfnamefont {G.}~\bibnamefont {Liu}}, \bibinfo
  {author} {\bibfnamefont {P.}~\bibnamefont {Peng}}, \bibinfo {author}
  {\bibfnamefont {B.}~\bibnamefont {Zhang}}, \bibinfo {author} {\bibfnamefont
  {S.}~\bibnamefont {Jandura}}, \bibinfo {author} {\bibfnamefont
  {J.}~\bibnamefont {Claes}}, \bibinfo {author} {\bibfnamefont {A.~P.}\
  \bibnamefont {Burgers}}, \bibinfo {author} {\bibfnamefont {G.}~\bibnamefont
  {Pupillo}}, \bibinfo {author} {\bibfnamefont {S.}~\bibnamefont {Puri}},\ and\
  \bibinfo {author} {\bibfnamefont {J.~D.}\ \bibnamefont {Thompson}},\
  }\bibfield  {title} {\bibinfo {title} {High-fidelity gates and mid-circuit
  erasure conversion in an atomic qubit},\ }\href
  {https://doi.org/10.1038/s41586-023-06438-1} {\bibfield  {journal} {\bibinfo
  {journal} {Nature}\ }\textbf {\bibinfo {volume} {622}},\ \bibinfo {pages}
  {279} (\bibinfo {year} {2023})}\BibitemShut {NoStop}%
\bibitem [{\citenamefont {Jenkins}\ \emph {et~al.}(2022)\citenamefont
  {Jenkins}, \citenamefont {Lis}, \citenamefont {Senoo}, \citenamefont
  {McGrew},\ and\ \citenamefont {Kaufman}}]{Jenkins2022}%
  \BibitemOpen
  \bibfield  {author} {\bibinfo {author} {\bibfnamefont {A.}~\bibnamefont
  {Jenkins}}, \bibinfo {author} {\bibfnamefont {J.~W.}\ \bibnamefont {Lis}},
  \bibinfo {author} {\bibfnamefont {A.}~\bibnamefont {Senoo}}, \bibinfo
  {author} {\bibfnamefont {W.~F.}\ \bibnamefont {McGrew}},\ and\ \bibinfo
  {author} {\bibfnamefont {A.~M.}\ \bibnamefont {Kaufman}},\ }\bibfield
  {title} {\bibinfo {title} {Ytterbium nuclear-spin qubits in an optical
  tweezer array},\ }\href {https://doi.org/10.1103/PhysRevX.12.021027}
  {\bibfield  {journal} {\bibinfo  {journal} {Phys. Rev. X}\ }\textbf {\bibinfo
  {volume} {12}},\ \bibinfo {pages} {021027} (\bibinfo {year}
  {2022})}\BibitemShut {NoStop}%
\bibitem [{\citenamefont {Lukin}\ \emph {et~al.}(2001)\citenamefont {Lukin},
  \citenamefont {Fleischhauer}, \citenamefont {Cote}, \citenamefont {Duan},
  \citenamefont {Jaksch}, \citenamefont {Cirac},\ and\ \citenamefont
  {Zoller}}]{Lukin2001}%
  \BibitemOpen
  \bibfield  {author} {\bibinfo {author} {\bibfnamefont {M.~D.}\ \bibnamefont
  {Lukin}}, \bibinfo {author} {\bibfnamefont {M.}~\bibnamefont {Fleischhauer}},
  \bibinfo {author} {\bibfnamefont {R.}~\bibnamefont {Cote}}, \bibinfo {author}
  {\bibfnamefont {L.~M.}\ \bibnamefont {Duan}}, \bibinfo {author}
  {\bibfnamefont {D.}~\bibnamefont {Jaksch}}, \bibinfo {author} {\bibfnamefont
  {J.~I.}\ \bibnamefont {Cirac}},\ and\ \bibinfo {author} {\bibfnamefont
  {P.}~\bibnamefont {Zoller}},\ }\bibfield  {title} {\bibinfo {title} {Dipole
  blockade and quantum information processing in mesoscopic atomic ensembles},\
  }\href {https://doi.org/10.1103/PhysRevLett.87.037901} {\bibfield  {journal}
  {\bibinfo  {journal} {Phys. Rev. Lett.}\ }\textbf {\bibinfo {volume} {87}},\
  \bibinfo {pages} {037901} (\bibinfo {year} {2001})}\BibitemShut {NoStop}%
\bibitem [{\citenamefont {Levine}\ \emph {et~al.}(2018)\citenamefont {Levine},
  \citenamefont {Keesling}, \citenamefont {Omran}, \citenamefont {Bernien},
  \citenamefont {Schwartz}, \citenamefont {Zibrov}, \citenamefont {Endres},
  \citenamefont {Greiner}, \citenamefont {Vuleti\ifmmode~\acute{c}\else
  \'{c}\fi{}},\ and\ \citenamefont {Lukin}}]{Levine2018}%
  \BibitemOpen
  \bibfield  {author} {\bibinfo {author} {\bibfnamefont {H.}~\bibnamefont
  {Levine}}, \bibinfo {author} {\bibfnamefont {A.}~\bibnamefont {Keesling}},
  \bibinfo {author} {\bibfnamefont {A.}~\bibnamefont {Omran}}, \bibinfo
  {author} {\bibfnamefont {H.}~\bibnamefont {Bernien}}, \bibinfo {author}
  {\bibfnamefont {S.}~\bibnamefont {Schwartz}}, \bibinfo {author}
  {\bibfnamefont {A.~S.}\ \bibnamefont {Zibrov}}, \bibinfo {author}
  {\bibfnamefont {M.}~\bibnamefont {Endres}}, \bibinfo {author} {\bibfnamefont
  {M.}~\bibnamefont {Greiner}}, \bibinfo {author} {\bibfnamefont
  {V.}~\bibnamefont {Vuleti\ifmmode~\acute{c}\else \'{c}\fi{}}},\ and\ \bibinfo
  {author} {\bibfnamefont {M.~D.}\ \bibnamefont {Lukin}},\ }\bibfield  {title}
  {\bibinfo {title} {High-fidelity control and entanglement of rydberg-atom
  qubits},\ }\href {https://doi.org/10.1103/PhysRevLett.121.123603} {\bibfield
  {journal} {\bibinfo  {journal} {Phys. Rev. Lett.}\ }\textbf {\bibinfo
  {volume} {121}},\ \bibinfo {pages} {123603} (\bibinfo {year}
  {2018})}\BibitemShut {NoStop}%
\bibitem [{\citenamefont {Levine}\ \emph {et~al.}(2019)\citenamefont {Levine},
  \citenamefont {Keesling}, \citenamefont {Semeghini}, \citenamefont {Omran},
  \citenamefont {Wang}, \citenamefont {Ebadi}, \citenamefont {Bernien},
  \citenamefont {Greiner}, \citenamefont {Vuleti\ifmmode~\acute{c}\else
  \'{c}\fi{}}, \citenamefont {Pichler},\ and\ \citenamefont
  {Lukin}}]{Levine2019}%
  \BibitemOpen
  \bibfield  {author} {\bibinfo {author} {\bibfnamefont {H.}~\bibnamefont
  {Levine}}, \bibinfo {author} {\bibfnamefont {A.}~\bibnamefont {Keesling}},
  \bibinfo {author} {\bibfnamefont {G.}~\bibnamefont {Semeghini}}, \bibinfo
  {author} {\bibfnamefont {A.}~\bibnamefont {Omran}}, \bibinfo {author}
  {\bibfnamefont {T.~T.}\ \bibnamefont {Wang}}, \bibinfo {author}
  {\bibfnamefont {S.}~\bibnamefont {Ebadi}}, \bibinfo {author} {\bibfnamefont
  {H.}~\bibnamefont {Bernien}}, \bibinfo {author} {\bibfnamefont
  {M.}~\bibnamefont {Greiner}}, \bibinfo {author} {\bibfnamefont
  {V.}~\bibnamefont {Vuleti\ifmmode~\acute{c}\else \'{c}\fi{}}}, \bibinfo
  {author} {\bibfnamefont {H.}~\bibnamefont {Pichler}},\ and\ \bibinfo {author}
  {\bibfnamefont {M.~D.}\ \bibnamefont {Lukin}},\ }\bibfield  {title} {\bibinfo
  {title} {Parallel implementation of high-fidelity multiqubit gates with
  neutral atoms},\ }\href {https://doi.org/10.1103/PhysRevLett.123.170503}
  {\bibfield  {journal} {\bibinfo  {journal} {Phys. Rev. Lett.}\ }\textbf
  {\bibinfo {volume} {123}},\ \bibinfo {pages} {170503} (\bibinfo {year}
  {2019})}\BibitemShut {NoStop}%
\bibitem [{\citenamefont {Ebadi}\ \emph {et~al.}(2021)\citenamefont {Ebadi},
  \citenamefont {Wang}, \citenamefont {Levine}, \citenamefont {Keesling},
  \citenamefont {Semeghini}, \citenamefont {Omran}, \citenamefont {Bluvstein},
  \citenamefont {Samajdar}, \citenamefont {Pichler}, \citenamefont {Ho},
  \citenamefont {Choi}, \citenamefont {Sachdev}, \citenamefont {Greiner},
  \citenamefont {Vuleti{\'{c}}},\ and\ \citenamefont {Lukin}}]{Ebadi2021}%
  \BibitemOpen
  \bibfield  {author} {\bibinfo {author} {\bibfnamefont {S.}~\bibnamefont
  {Ebadi}}, \bibinfo {author} {\bibfnamefont {T.~T.}\ \bibnamefont {Wang}},
  \bibinfo {author} {\bibfnamefont {H.}~\bibnamefont {Levine}}, \bibinfo
  {author} {\bibfnamefont {A.}~\bibnamefont {Keesling}}, \bibinfo {author}
  {\bibfnamefont {G.}~\bibnamefont {Semeghini}}, \bibinfo {author}
  {\bibfnamefont {A.}~\bibnamefont {Omran}}, \bibinfo {author} {\bibfnamefont
  {D.}~\bibnamefont {Bluvstein}}, \bibinfo {author} {\bibfnamefont
  {R.}~\bibnamefont {Samajdar}}, \bibinfo {author} {\bibfnamefont
  {H.}~\bibnamefont {Pichler}}, \bibinfo {author} {\bibfnamefont {W.~W.}\
  \bibnamefont {Ho}}, \bibinfo {author} {\bibfnamefont {S.}~\bibnamefont
  {Choi}}, \bibinfo {author} {\bibfnamefont {S.}~\bibnamefont {Sachdev}},
  \bibinfo {author} {\bibfnamefont {M.}~\bibnamefont {Greiner}}, \bibinfo
  {author} {\bibfnamefont {V.}~\bibnamefont {Vuleti{\'{c}}}},\ and\ \bibinfo
  {author} {\bibfnamefont {M.~D.}\ \bibnamefont {Lukin}},\ }\bibfield  {title}
  {\bibinfo {title} {Quantum phases of matter on a 256-atom programmable
  quantum simulator},\ }\href {https://doi.org/10.1038/s41586-021-03582-4}
  {\bibfield  {journal} {\bibinfo  {journal} {Nature}\ }\textbf {\bibinfo
  {volume} {595}},\ \bibinfo {pages} {227} (\bibinfo {year}
  {2021})}\BibitemShut {NoStop}%
\bibitem [{\citenamefont {Fu}\ \emph {et~al.}(2022)\citenamefont {Fu},
  \citenamefont {Xu}, \citenamefont {Sun}, \citenamefont {Liu}, \citenamefont
  {He}, \citenamefont {Li}, \citenamefont {Liu}, \citenamefont {Li},
  \citenamefont {Wang}, \citenamefont {Liu},\ and\ \citenamefont
  {Zhan}}]{Fu2022}%
  \BibitemOpen
  \bibfield  {author} {\bibinfo {author} {\bibfnamefont {Z.}~\bibnamefont
  {Fu}}, \bibinfo {author} {\bibfnamefont {P.}~\bibnamefont {Xu}}, \bibinfo
  {author} {\bibfnamefont {Y.}~\bibnamefont {Sun}}, \bibinfo {author}
  {\bibfnamefont {Y.-Y.}\ \bibnamefont {Liu}}, \bibinfo {author} {\bibfnamefont
  {X.-D.}\ \bibnamefont {He}}, \bibinfo {author} {\bibfnamefont
  {X.}~\bibnamefont {Li}}, \bibinfo {author} {\bibfnamefont {M.}~\bibnamefont
  {Liu}}, \bibinfo {author} {\bibfnamefont {R.-B.}\ \bibnamefont {Li}},
  \bibinfo {author} {\bibfnamefont {J.}~\bibnamefont {Wang}}, \bibinfo {author}
  {\bibfnamefont {L.}~\bibnamefont {Liu}},\ and\ \bibinfo {author}
  {\bibfnamefont {M.-S.}\ \bibnamefont {Zhan}},\ }\bibfield  {title} {\bibinfo
  {title} {High-fidelity entanglement of neutral atoms via a rydberg-mediated
  single-modulated-pulse controlled-phase gate},\ }\href
  {https://doi.org/10.1103/PhysRevA.105.042430} {\bibfield  {journal} {\bibinfo
   {journal} {Phys. Rev. A}\ }\textbf {\bibinfo {volume} {105}},\ \bibinfo
  {pages} {042430} (\bibinfo {year} {2022})}\BibitemShut {NoStop}%
\bibitem [{\citenamefont {Evered}\ \emph {et~al.}(2023)\citenamefont {Evered},
  \citenamefont {Bluvstein}, \citenamefont {Kalinowski}, \citenamefont {Ebadi},
  \citenamefont {Manovitz}, \citenamefont {Zhou}, \citenamefont {Li},
  \citenamefont {Geim}, \citenamefont {Wang}, \citenamefont {Maskara},
  \citenamefont {Levine}, \citenamefont {Semeghini}, \citenamefont {Greiner},
  \citenamefont {Vuleti{\'{c}}},\ and\ \citenamefont {Lukin}}]{Evered2023}%
  \BibitemOpen
  \bibfield  {author} {\bibinfo {author} {\bibfnamefont {S.~J.}\ \bibnamefont
  {Evered}}, \bibinfo {author} {\bibfnamefont {D.}~\bibnamefont {Bluvstein}},
  \bibinfo {author} {\bibfnamefont {M.}~\bibnamefont {Kalinowski}}, \bibinfo
  {author} {\bibfnamefont {S.}~\bibnamefont {Ebadi}}, \bibinfo {author}
  {\bibfnamefont {T.}~\bibnamefont {Manovitz}}, \bibinfo {author}
  {\bibfnamefont {H.}~\bibnamefont {Zhou}}, \bibinfo {author} {\bibfnamefont
  {S.~H.}\ \bibnamefont {Li}}, \bibinfo {author} {\bibfnamefont {A.~A.}\
  \bibnamefont {Geim}}, \bibinfo {author} {\bibfnamefont {T.~T.}\ \bibnamefont
  {Wang}}, \bibinfo {author} {\bibfnamefont {N.}~\bibnamefont {Maskara}},
  \bibinfo {author} {\bibfnamefont {H.}~\bibnamefont {Levine}}, \bibinfo
  {author} {\bibfnamefont {G.}~\bibnamefont {Semeghini}}, \bibinfo {author}
  {\bibfnamefont {M.}~\bibnamefont {Greiner}}, \bibinfo {author} {\bibfnamefont
  {V.}~\bibnamefont {Vuleti{\'{c}}}},\ and\ \bibinfo {author} {\bibfnamefont
  {M.~D.}\ \bibnamefont {Lukin}},\ }\bibfield  {title} {\bibinfo {title}
  {High-fidelity parallel entangling gates on a neutral-atom quantum
  computer},\ }\href {https://doi.org/10.1038/s41586-023-06481-y} {\bibfield
  {journal} {\bibinfo  {journal} {Nature}\ }\textbf {\bibinfo {volume} {622}},\
  \bibinfo {pages} {268} (\bibinfo {year} {2023})}\BibitemShut {NoStop}%
\bibitem [{\citenamefont {Bluvstein}\ \emph {et~al.}(2024)\citenamefont
  {Bluvstein}, \citenamefont {Evered}, \citenamefont {Geim}, \citenamefont
  {Li}, \citenamefont {Zhou}, \citenamefont {Manovitz}, \citenamefont {Ebadi},
  \citenamefont {Cain}, \citenamefont {Kalinowski}, \citenamefont {Hangleiter},
  \citenamefont {Bonilla~Ataides}, \citenamefont {Maskara}, \citenamefont
  {Cong}, \citenamefont {Gao}, \citenamefont {Sales~Rodriguez}, \citenamefont
  {Karolyshyn}, \citenamefont {Semeghini}, \citenamefont {Gullans},
  \citenamefont {Greiner}, \citenamefont {Vuleti{\'{c}}},\ and\ \citenamefont
  {Lukin}}]{Bluvstein2023}%
  \BibitemOpen
  \bibfield  {author} {\bibinfo {author} {\bibfnamefont {D.}~\bibnamefont
  {Bluvstein}}, \bibinfo {author} {\bibfnamefont {S.~J.}\ \bibnamefont
  {Evered}}, \bibinfo {author} {\bibfnamefont {A.~A.}\ \bibnamefont {Geim}},
  \bibinfo {author} {\bibfnamefont {S.~H.}\ \bibnamefont {Li}}, \bibinfo
  {author} {\bibfnamefont {H.}~\bibnamefont {Zhou}}, \bibinfo {author}
  {\bibfnamefont {T.}~\bibnamefont {Manovitz}}, \bibinfo {author}
  {\bibfnamefont {S.}~\bibnamefont {Ebadi}}, \bibinfo {author} {\bibfnamefont
  {M.}~\bibnamefont {Cain}}, \bibinfo {author} {\bibfnamefont {M.}~\bibnamefont
  {Kalinowski}}, \bibinfo {author} {\bibfnamefont {D.}~\bibnamefont
  {Hangleiter}}, \bibinfo {author} {\bibfnamefont {J.~P.}\ \bibnamefont
  {Bonilla~Ataides}}, \bibinfo {author} {\bibfnamefont {N.}~\bibnamefont
  {Maskara}}, \bibinfo {author} {\bibfnamefont {I.}~\bibnamefont {Cong}},
  \bibinfo {author} {\bibfnamefont {X.}~\bibnamefont {Gao}}, \bibinfo {author}
  {\bibfnamefont {P.}~\bibnamefont {Sales~Rodriguez}}, \bibinfo {author}
  {\bibfnamefont {T.}~\bibnamefont {Karolyshyn}}, \bibinfo {author}
  {\bibfnamefont {G.}~\bibnamefont {Semeghini}}, \bibinfo {author}
  {\bibfnamefont {M.~J.}\ \bibnamefont {Gullans}}, \bibinfo {author}
  {\bibfnamefont {M.}~\bibnamefont {Greiner}}, \bibinfo {author} {\bibfnamefont
  {V.}~\bibnamefont {Vuleti{\'{c}}}},\ and\ \bibinfo {author} {\bibfnamefont
  {M.~D.}\ \bibnamefont {Lukin}},\ }\bibfield  {title} {\bibinfo {title}
  {Logical quantum processor based on reconfigurable atom arrays},\ }\href
  {https://doi.org/10.1038/s41586-023-06927-3} {\bibfield  {journal} {\bibinfo
  {journal} {Nature}\ }\textbf {\bibinfo {volume} {626}},\ \bibinfo {pages}
  {58} (\bibinfo {year} {2024})}\BibitemShut {NoStop}%
\bibitem [{\citenamefont {Radnaev}\ \emph {et~al.}(2024)\citenamefont
  {Radnaev}, \citenamefont {Chung}, \citenamefont {Cole}, \citenamefont
  {Mason}, \citenamefont {Ballance}, \citenamefont {Bedalov}, \citenamefont
  {Belknap}, \citenamefont {Berman}, \citenamefont {Blakely}, \citenamefont
  {Bloomfield}, \citenamefont {Buttler}, \citenamefont {Campbell},
  \citenamefont {Chopinaud}, \citenamefont {Copenhaver}, \citenamefont {Dawes},
  \citenamefont {Eubanks}, \citenamefont {Friss}, \citenamefont {Garcia},
  \citenamefont {Gilbert}, \citenamefont {Gillette}, \citenamefont {Goiporia},
  \citenamefont {Gokhale}, \citenamefont {Goldwin}, \citenamefont {Goodwin},
  \citenamefont {Graham}, \citenamefont {Guttormsson}, \citenamefont {Hickman},
  \citenamefont {Hurtley}, \citenamefont {Iliev}, \citenamefont {Jones},
  \citenamefont {Jones}, \citenamefont {Kuper}, \citenamefont {Lewis},
  \citenamefont {Lichtman}, \citenamefont {Majdeteimouri}, \citenamefont
  {Mason}, \citenamefont {McMaster}, \citenamefont {Miles}, \citenamefont
  {Mitchell}, \citenamefont {Murphree}, \citenamefont {Neff-Mallon},
  \citenamefont {Oh}, \citenamefont {Omole}, \citenamefont {Simon},
  \citenamefont {Pederson}, \citenamefont {Perlin}, \citenamefont {Reiter},
  \citenamefont {Rines}, \citenamefont {Romlow}, \citenamefont {Scott},
  \citenamefont {Stiefvater}, \citenamefont {Tanner}, \citenamefont {Tucker},
  \citenamefont {Vinogradov}, \citenamefont {Warter}, \citenamefont {Yeo},
  \citenamefont {Saffman},\ and\ \citenamefont {Noel}}]{Radnaev2024}%
  \BibitemOpen
  \bibfield  {author} {\bibinfo {author} {\bibfnamefont {A.~G.}\ \bibnamefont
  {Radnaev}}, \bibinfo {author} {\bibfnamefont {W.~C.}\ \bibnamefont {Chung}},
  \bibinfo {author} {\bibfnamefont {D.~C.}\ \bibnamefont {Cole}}, \bibinfo
  {author} {\bibfnamefont {D.}~\bibnamefont {Mason}}, \bibinfo {author}
  {\bibfnamefont {T.~G.}\ \bibnamefont {Ballance}}, \bibinfo {author}
  {\bibfnamefont {M.~J.}\ \bibnamefont {Bedalov}}, \bibinfo {author}
  {\bibfnamefont {D.~A.}\ \bibnamefont {Belknap}}, \bibinfo {author}
  {\bibfnamefont {M.~R.}\ \bibnamefont {Berman}}, \bibinfo {author}
  {\bibfnamefont {M.}~\bibnamefont {Blakely}}, \bibinfo {author} {\bibfnamefont
  {I.~L.}\ \bibnamefont {Bloomfield}}, \bibinfo {author} {\bibfnamefont
  {P.~D.}\ \bibnamefont {Buttler}}, \bibinfo {author} {\bibfnamefont
  {C.}~\bibnamefont {Campbell}}, \bibinfo {author} {\bibfnamefont
  {A.}~\bibnamefont {Chopinaud}}, \bibinfo {author} {\bibfnamefont
  {E.}~\bibnamefont {Copenhaver}}, \bibinfo {author} {\bibfnamefont {M.~K.}\
  \bibnamefont {Dawes}}, \bibinfo {author} {\bibfnamefont {S.~Y.}\ \bibnamefont
  {Eubanks}}, \bibinfo {author} {\bibfnamefont {A.~J.}\ \bibnamefont {Friss}},
  \bibinfo {author} {\bibfnamefont {D.~M.}\ \bibnamefont {Garcia}}, \bibinfo
  {author} {\bibfnamefont {J.}~\bibnamefont {Gilbert}}, \bibinfo {author}
  {\bibfnamefont {M.}~\bibnamefont {Gillette}}, \bibinfo {author}
  {\bibfnamefont {P.}~\bibnamefont {Goiporia}}, \bibinfo {author}
  {\bibfnamefont {P.}~\bibnamefont {Gokhale}}, \bibinfo {author} {\bibfnamefont
  {J.}~\bibnamefont {Goldwin}}, \bibinfo {author} {\bibfnamefont
  {D.}~\bibnamefont {Goodwin}}, \bibinfo {author} {\bibfnamefont {T.~M.}\
  \bibnamefont {Graham}}, \bibinfo {author} {\bibfnamefont {C.}~\bibnamefont
  {Guttormsson}}, \bibinfo {author} {\bibfnamefont {G.~T.}\ \bibnamefont
  {Hickman}}, \bibinfo {author} {\bibfnamefont {L.}~\bibnamefont {Hurtley}},
  \bibinfo {author} {\bibfnamefont {M.}~\bibnamefont {Iliev}}, \bibinfo
  {author} {\bibfnamefont {E.~B.}\ \bibnamefont {Jones}}, \bibinfo {author}
  {\bibfnamefont {R.~A.}\ \bibnamefont {Jones}}, \bibinfo {author}
  {\bibfnamefont {K.~W.}\ \bibnamefont {Kuper}}, \bibinfo {author}
  {\bibfnamefont {T.~B.}\ \bibnamefont {Lewis}}, \bibinfo {author}
  {\bibfnamefont {M.~T.}\ \bibnamefont {Lichtman}}, \bibinfo {author}
  {\bibfnamefont {F.}~\bibnamefont {Majdeteimouri}}, \bibinfo {author}
  {\bibfnamefont {J.~J.}\ \bibnamefont {Mason}}, \bibinfo {author}
  {\bibfnamefont {J.~K.}\ \bibnamefont {McMaster}}, \bibinfo {author}
  {\bibfnamefont {J.~A.}\ \bibnamefont {Miles}}, \bibinfo {author}
  {\bibfnamefont {P.~T.}\ \bibnamefont {Mitchell}}, \bibinfo {author}
  {\bibfnamefont {J.~D.}\ \bibnamefont {Murphree}}, \bibinfo {author}
  {\bibfnamefont {N.~A.}\ \bibnamefont {Neff-Mallon}}, \bibinfo {author}
  {\bibfnamefont {T.}~\bibnamefont {Oh}}, \bibinfo {author} {\bibfnamefont
  {V.}~\bibnamefont {Omole}}, \bibinfo {author} {\bibfnamefont {C.~P.}\
  \bibnamefont {Simon}}, \bibinfo {author} {\bibfnamefont {N.}~\bibnamefont
  {Pederson}}, \bibinfo {author} {\bibfnamefont {M.~A.}\ \bibnamefont
  {Perlin}}, \bibinfo {author} {\bibfnamefont {A.}~\bibnamefont {Reiter}},
  \bibinfo {author} {\bibfnamefont {R.}~\bibnamefont {Rines}}, \bibinfo
  {author} {\bibfnamefont {P.}~\bibnamefont {Romlow}}, \bibinfo {author}
  {\bibfnamefont {A.~M.}\ \bibnamefont {Scott}}, \bibinfo {author}
  {\bibfnamefont {D.}~\bibnamefont {Stiefvater}}, \bibinfo {author}
  {\bibfnamefont {J.~R.}\ \bibnamefont {Tanner}}, \bibinfo {author}
  {\bibfnamefont {A.~K.}\ \bibnamefont {Tucker}}, \bibinfo {author}
  {\bibfnamefont {I.~V.}\ \bibnamefont {Vinogradov}}, \bibinfo {author}
  {\bibfnamefont {M.~L.}\ \bibnamefont {Warter}}, \bibinfo {author}
  {\bibfnamefont {M.}~\bibnamefont {Yeo}}, \bibinfo {author} {\bibfnamefont
  {M.}~\bibnamefont {Saffman}},\ and\ \bibinfo {author} {\bibfnamefont {T.~W.}\
  \bibnamefont {Noel}},\ }\href {https://arxiv.org/abs/2408.08288} {\bibinfo
  {title} {A universal neutral-atom quantum computer with individual optical
  addressing and non-destructive readout}} (\bibinfo {year} {2024}),\ \Eprint
  {https://arxiv.org/abs/2408.08288} {arXiv:2408.08288 [quant-ph]} \BibitemShut
  {NoStop}%
\bibitem [{\citenamefont {Tsai}\ \emph {et~al.}(2024)\citenamefont {Tsai},
  \citenamefont {Sun}, \citenamefont {Shaw}, \citenamefont {Finkelstein},\ and\
  \citenamefont {Endres}}]{Tsai2024}%
  \BibitemOpen
  \bibfield  {author} {\bibinfo {author} {\bibfnamefont {R.~B.-S.}\
  \bibnamefont {Tsai}}, \bibinfo {author} {\bibfnamefont {X.}~\bibnamefont
  {Sun}}, \bibinfo {author} {\bibfnamefont {A.~L.}\ \bibnamefont {Shaw}},
  \bibinfo {author} {\bibfnamefont {R.}~\bibnamefont {Finkelstein}},\ and\
  \bibinfo {author} {\bibfnamefont {M.}~\bibnamefont {Endres}},\ }\href
  {https://arxiv.org/abs/2407.20184} {\bibinfo {title} {Benchmarking and linear
  response modeling of high-fidelity rydberg gates}} (\bibinfo {year} {2024}),\
  \Eprint {https://arxiv.org/abs/2407.20184} {arXiv:2407.20184 [quant-ph]}
  \BibitemShut {NoStop}%
\bibitem [{\citenamefont {Huang}\ \emph {et~al.}(2020)\citenamefont {Huang},
  \citenamefont {Wu}, \citenamefont {Fan},\ and\ \citenamefont
  {Zhu}}]{Huang2020}%
  \BibitemOpen
  \bibfield  {author} {\bibinfo {author} {\bibfnamefont {H.-L.}\ \bibnamefont
  {Huang}}, \bibinfo {author} {\bibfnamefont {D.}~\bibnamefont {Wu}}, \bibinfo
  {author} {\bibfnamefont {D.}~\bibnamefont {Fan}},\ and\ \bibinfo {author}
  {\bibfnamefont {X.}~\bibnamefont {Zhu}},\ }\bibfield  {title} {\bibinfo
  {title} {Superconducting quantum computing: a review},\ }\href
  {https://doi.org/10.1007/s11432-020-2881-9} {\bibfield  {journal} {\bibinfo
  {journal} {Science China Information Sciences}\ }\textbf {\bibinfo {volume}
  {63}},\ \bibinfo {pages} {180501} (\bibinfo {year} {2020})}\BibitemShut
  {NoStop}%
\bibitem [{\citenamefont {Bruzewicz}\ \emph {et~al.}(2019)\citenamefont
  {Bruzewicz}, \citenamefont {Chiaverini}, \citenamefont {McConnell},\ and\
  \citenamefont {Sage}}]{Bruzewicz2019}%
  \BibitemOpen
  \bibfield  {author} {\bibinfo {author} {\bibfnamefont {C.~D.}\ \bibnamefont
  {Bruzewicz}}, \bibinfo {author} {\bibfnamefont {J.}~\bibnamefont
  {Chiaverini}}, \bibinfo {author} {\bibfnamefont {R.}~\bibnamefont
  {McConnell}},\ and\ \bibinfo {author} {\bibfnamefont {J.~M.}\ \bibnamefont
  {Sage}},\ }\bibfield  {title} {\bibinfo {title} {Trapped-ion quantum
  computing: Progress and challenges},\ }\href
  {https://doi.org/10.1063/1.5088164} {\bibfield  {journal} {\bibinfo
  {journal} {Applied Physics Reviews}\ }\textbf {\bibinfo {volume} {6}},\
  \bibinfo {pages} {021314} (\bibinfo {year} {2019})}\BibitemShut {NoStop}%
\bibitem [{\citenamefont {Graham}\ \emph {et~al.}(2022)\citenamefont {Graham},
  \citenamefont {Song}, \citenamefont {Scott}, \citenamefont {Poole},
  \citenamefont {Phuttitarn}, \citenamefont {Jooya}, \citenamefont {Eichler},
  \citenamefont {Jiang}, \citenamefont {Marra}, \citenamefont {Grinkemeyer},
  \citenamefont {Kwon}, \citenamefont {Ebert}, \citenamefont {Cherek},
  \citenamefont {Lichtman}, \citenamefont {Gillette}, \citenamefont {Gilbert},
  \citenamefont {Bowman}, \citenamefont {Ballance}, \citenamefont {Campbell},
  \citenamefont {Dahl}, \citenamefont {Crawford}, \citenamefont {Blunt},
  \citenamefont {Rogers}, \citenamefont {Noel},\ and\ \citenamefont
  {Saffman}}]{Graham2022}%
  \BibitemOpen
  \bibfield  {author} {\bibinfo {author} {\bibfnamefont {T.~M.}\ \bibnamefont
  {Graham}}, \bibinfo {author} {\bibfnamefont {Y.}~\bibnamefont {Song}},
  \bibinfo {author} {\bibfnamefont {J.}~\bibnamefont {Scott}}, \bibinfo
  {author} {\bibfnamefont {C.}~\bibnamefont {Poole}}, \bibinfo {author}
  {\bibfnamefont {L.}~\bibnamefont {Phuttitarn}}, \bibinfo {author}
  {\bibfnamefont {K.}~\bibnamefont {Jooya}}, \bibinfo {author} {\bibfnamefont
  {P.}~\bibnamefont {Eichler}}, \bibinfo {author} {\bibfnamefont
  {X.}~\bibnamefont {Jiang}}, \bibinfo {author} {\bibfnamefont
  {A.}~\bibnamefont {Marra}}, \bibinfo {author} {\bibfnamefont
  {B.}~\bibnamefont {Grinkemeyer}}, \bibinfo {author} {\bibfnamefont
  {M.}~\bibnamefont {Kwon}}, \bibinfo {author} {\bibfnamefont {M.}~\bibnamefont
  {Ebert}}, \bibinfo {author} {\bibfnamefont {J.}~\bibnamefont {Cherek}},
  \bibinfo {author} {\bibfnamefont {M.~T.}\ \bibnamefont {Lichtman}}, \bibinfo
  {author} {\bibfnamefont {M.}~\bibnamefont {Gillette}}, \bibinfo {author}
  {\bibfnamefont {J.}~\bibnamefont {Gilbert}}, \bibinfo {author} {\bibfnamefont
  {D.}~\bibnamefont {Bowman}}, \bibinfo {author} {\bibfnamefont
  {T.}~\bibnamefont {Ballance}}, \bibinfo {author} {\bibfnamefont
  {C.}~\bibnamefont {Campbell}}, \bibinfo {author} {\bibfnamefont {E.~D.}\
  \bibnamefont {Dahl}}, \bibinfo {author} {\bibfnamefont {O.}~\bibnamefont
  {Crawford}}, \bibinfo {author} {\bibfnamefont {N.~S.}\ \bibnamefont {Blunt}},
  \bibinfo {author} {\bibfnamefont {B.}~\bibnamefont {Rogers}}, \bibinfo
  {author} {\bibfnamefont {T.}~\bibnamefont {Noel}},\ and\ \bibinfo {author}
  {\bibfnamefont {M.}~\bibnamefont {Saffman}},\ }\bibfield  {title} {\bibinfo
  {title} {Multi-qubit entanglement and algorithms on a neutral-atom quantum
  computer},\ }\href {https://doi.org/10.1038/s41586-022-04603-6} {\bibfield
  {journal} {\bibinfo  {journal} {Nature}\ }\textbf {\bibinfo {volume} {604}},\
  \bibinfo {pages} {457} (\bibinfo {year} {2022})}\BibitemShut {NoStop}%
\bibitem [{\citenamefont {Tian}\ \emph {et~al.}(2024)\citenamefont {Tian},
  \citenamefont {Chang}, \citenamefont {Lv}, \citenamefont {Yang},
  \citenamefont {Wang}, \citenamefont {Yang}, \citenamefont {Zhang},
  \citenamefont {Li},\ and\ \citenamefont {Zhang}}]{Tian2024}%
  \BibitemOpen
  \bibfield  {author} {\bibinfo {author} {\bibfnamefont {Z.}~\bibnamefont
  {Tian}}, \bibinfo {author} {\bibfnamefont {H.}~\bibnamefont {Chang}},
  \bibinfo {author} {\bibfnamefont {X.}~\bibnamefont {Lv}}, \bibinfo {author}
  {\bibfnamefont {M.}~\bibnamefont {Yang}}, \bibinfo {author} {\bibfnamefont
  {Z.}~\bibnamefont {Wang}}, \bibinfo {author} {\bibfnamefont {P.}~\bibnamefont
  {Yang}}, \bibinfo {author} {\bibfnamefont {P.}~\bibnamefont {Zhang}},
  \bibinfo {author} {\bibfnamefont {G.}~\bibnamefont {Li}},\ and\ \bibinfo
  {author} {\bibfnamefont {T.}~\bibnamefont {Zhang}},\ }\bibfield  {title}
  {\bibinfo {title} {Resolved raman sideband cooling of a single optically
  trapped cesium atom},\ }\href {https://doi.org/10.1364/OL.514160} {\bibfield
  {journal} {\bibinfo  {journal} {Opt. Lett.}\ }\textbf {\bibinfo {volume}
  {49}},\ \bibinfo {pages} {542} (\bibinfo {year} {2024})}\BibitemShut
  {NoStop}%
\bibitem [{\citenamefont {Ryabtsev}\ \emph {et~al.}(2011)\citenamefont
  {Ryabtsev}, \citenamefont {Beterov}, \citenamefont {Tretyakov}, \citenamefont
  {Entin},\ and\ \citenamefont {Yakshina}}]{Ryabtsev2011}%
  \BibitemOpen
  \bibfield  {author} {\bibinfo {author} {\bibfnamefont {I.~I.}\ \bibnamefont
  {Ryabtsev}}, \bibinfo {author} {\bibfnamefont {I.~I.}\ \bibnamefont
  {Beterov}}, \bibinfo {author} {\bibfnamefont {D.~B.}\ \bibnamefont
  {Tretyakov}}, \bibinfo {author} {\bibfnamefont {V.~M.}\ \bibnamefont
  {Entin}},\ and\ \bibinfo {author} {\bibfnamefont {E.~A.}\ \bibnamefont
  {Yakshina}},\ }\bibfield  {title} {\bibinfo {title} {Doppler- and recoil-free
  laser excitation of rydberg states via three-photon transitions},\ }\href
  {https://doi.org/10.1103/PhysRevA.84.053409} {\bibfield  {journal} {\bibinfo
  {journal} {Phys. Rev. A}\ }\textbf {\bibinfo {volume} {84}},\ \bibinfo
  {pages} {053409} (\bibinfo {year} {2011})}\BibitemShut {NoStop}%
\bibitem [{\citenamefont {Entin}\ \emph {et~al.}(2013)\citenamefont {Entin},
  \citenamefont {Yakshina}, \citenamefont {Tretyakov}, \citenamefont
  {Beterov},\ and\ \citenamefont {Ryabtsev}}]{Entin2013}%
  \BibitemOpen
  \bibfield  {author} {\bibinfo {author} {\bibfnamefont {V.~M.}\ \bibnamefont
  {Entin}}, \bibinfo {author} {\bibfnamefont {E.~A.}\ \bibnamefont {Yakshina}},
  \bibinfo {author} {\bibfnamefont {D.~B.}\ \bibnamefont {Tretyakov}}, \bibinfo
  {author} {\bibfnamefont {I.~I.}\ \bibnamefont {Beterov}},\ and\ \bibinfo
  {author} {\bibfnamefont {I.~I.}\ \bibnamefont {Ryabtsev}},\ }\bibfield
  {title} {\bibinfo {title} {Spectroscopy of the three-photon laser excitation
  of cold rubidium rydberg atoms in a magneto-optical trap},\ }\href
  {https://doi.org/10.1134/S1063776113040110} {\bibfield  {journal} {\bibinfo
  {journal} {Journal of Experimental and Theoretical Physics}\ }\textbf
  {\bibinfo {volume} {116}},\ \bibinfo {pages} {721} (\bibinfo {year}
  {2013})}\BibitemShut {NoStop}%
\bibitem [{\citenamefont {Yakshina}\ \emph {et~al.}(2018)\citenamefont
  {Yakshina}, \citenamefont {Tretyakov}, \citenamefont {Entin}, \citenamefont
  {Beterov},\ and\ \citenamefont {Ryabtsev}}]{Yakshina2018}%
  \BibitemOpen
  \bibfield  {author} {\bibinfo {author} {\bibfnamefont {E.}~\bibnamefont
  {Yakshina}}, \bibinfo {author} {\bibfnamefont {D.}~\bibnamefont {Tretyakov}},
  \bibinfo {author} {\bibfnamefont {V.}~\bibnamefont {Entin}}, \bibinfo
  {author} {\bibfnamefont {I.}~\bibnamefont {Beterov}},\ and\ \bibinfo {author}
  {\bibfnamefont {I.}~\bibnamefont {Ryabtsev}},\ }\bibfield  {title} {\bibinfo
  {title} {Three-photon laser excitation of mesoscopic ensembles of cold
  rubidium rydberg atoms},\ }\href {https://doi.org/10.1070/QEL16765}
  {\bibfield  {journal} {\bibinfo  {journal} {Quantum Electronics}\ }\textbf
  {\bibinfo {volume} {48}},\ \bibinfo {pages} {886} (\bibinfo {year}
  {2018})}\BibitemShut {NoStop}%
\bibitem [{\citenamefont {Yakshina}\ \emph {et~al.}(2020)\citenamefont
  {Yakshina}, \citenamefont {Tretyakov}, \citenamefont {Entin}, \citenamefont
  {Beterov},\ and\ \citenamefont {Ryabtsev}}]{Yakshina2020}%
  \BibitemOpen
  \bibfield  {author} {\bibinfo {author} {\bibfnamefont {E.~A.}\ \bibnamefont
  {Yakshina}}, \bibinfo {author} {\bibfnamefont {D.~B.}\ \bibnamefont
  {Tretyakov}}, \bibinfo {author} {\bibfnamefont {V.~M.}\ \bibnamefont
  {Entin}}, \bibinfo {author} {\bibfnamefont {I.~I.}\ \bibnamefont {Beterov}},\
  and\ \bibinfo {author} {\bibfnamefont {I.~I.}\ \bibnamefont {Ryabtsev}},\
  }\bibfield  {title} {\bibinfo {title} {Observation of the dipole blockade
  effect in detecting rydberg atoms by the selective field ionization method},\
  }\href {https://doi.org/10.1134/S1063776120010215} {\bibfield  {journal}
  {\bibinfo  {journal} {Journal of Experimental and Theoretical Physics}\
  }\textbf {\bibinfo {volume} {130}},\ \bibinfo {pages} {170} (\bibinfo {year}
  {2020})}\BibitemShut {NoStop}%
\bibitem [{\citenamefont {Tretyakov}\ \emph {et~al.}(2022)\citenamefont
  {Tretyakov}, \citenamefont {Entin}, \citenamefont {Yakshina}, \citenamefont
  {Beterov},\ and\ \citenamefont {Ryabtsev}}]{Tretyakov2022}%
  \BibitemOpen
  \bibfield  {author} {\bibinfo {author} {\bibfnamefont {D.}~\bibnamefont
  {Tretyakov}}, \bibinfo {author} {\bibfnamefont {V.}~\bibnamefont {Entin}},
  \bibinfo {author} {\bibfnamefont {E.}~\bibnamefont {Yakshina}}, \bibinfo
  {author} {\bibfnamefont {I.}~\bibnamefont {Beterov}},\ and\ \bibinfo {author}
  {\bibfnamefont {I.}~\bibnamefont {Ryabtsev}},\ }\bibfield  {title} {\bibinfo
  {title} {Dynamics of three-photon laser excitation of mesoscopic ensembles of
  cold rubidium atoms to rydberg states},\ }\href
  {https://doi.org/10.1070/QEL18064} {\bibfield  {journal} {\bibinfo  {journal}
  {Quantum Electronics}\ }\textbf {\bibinfo {volume} {52}},\ \bibinfo {pages}
  {513} (\bibinfo {year} {2022})}\BibitemShut {NoStop}%
\bibitem [{\citenamefont {Beterov}\ \emph {et~al.}(2023)\citenamefont
  {Beterov}, \citenamefont {Yakshina}, \citenamefont {Tret'yakov},
  \citenamefont {Al'yanova}, \citenamefont {Skvortsova}, \citenamefont
  {Suliman}, \citenamefont {Zagirov}, \citenamefont {Entin},\ and\
  \citenamefont {Ryabtsev}}]{Beterov2023}%
  \BibitemOpen
  \bibfield  {author} {\bibinfo {author} {\bibfnamefont {I.~I.}\ \bibnamefont
  {Beterov}}, \bibinfo {author} {\bibfnamefont {E.~A.}\ \bibnamefont
  {Yakshina}}, \bibinfo {author} {\bibfnamefont {D.~B.}\ \bibnamefont
  {Tret'yakov}}, \bibinfo {author} {\bibfnamefont {N.~V.}\ \bibnamefont
  {Al'yanova}}, \bibinfo {author} {\bibfnamefont {D.~A.}\ \bibnamefont
  {Skvortsova}}, \bibinfo {author} {\bibfnamefont {G.}~\bibnamefont {Suliman}},
  \bibinfo {author} {\bibfnamefont {T.~R.}\ \bibnamefont {Zagirov}}, \bibinfo
  {author} {\bibfnamefont {V.~M.}\ \bibnamefont {Entin}},\ and\ \bibinfo
  {author} {\bibfnamefont {I.~I.}\ \bibnamefont {Ryabtsev}},\ }\bibfield
  {title} {\bibinfo {title} {Three-photon laser excitation of single rydberg
  rubidium atoms in an optical dipole trap},\ }\href
  {https://doi.org/10.1134/S1063776123080101} {\bibfield  {journal} {\bibinfo
  {journal} {Journal of Experimental and Theoretical Physics}\ }\textbf
  {\bibinfo {volume} {137}},\ \bibinfo {pages} {246} (\bibinfo {year}
  {2023})}\BibitemShut {NoStop}%
\bibitem [{\citenamefont {Beterov}\ \emph {et~al.}(2024)\citenamefont
  {Beterov}, \citenamefont {Yakshina}, \citenamefont {Suliman}, \citenamefont
  {Betleni}, \citenamefont {Prilutskaya}, \citenamefont {Skvortsova},
  \citenamefont {Zagirov}, \citenamefont {Tretyakov}, \citenamefont {Entin},
  \citenamefont {Bezuglov},\ and\ \citenamefont {Ryabtsev}}]{Beterov2024}%
  \BibitemOpen
  \bibfield  {author} {\bibinfo {author} {\bibfnamefont {I.~I.}\ \bibnamefont
  {Beterov}}, \bibinfo {author} {\bibfnamefont {E.~A.}\ \bibnamefont
  {Yakshina}}, \bibinfo {author} {\bibfnamefont {G.}~\bibnamefont {Suliman}},
  \bibinfo {author} {\bibfnamefont {P.~I.}\ \bibnamefont {Betleni}}, \bibinfo
  {author} {\bibfnamefont {A.~A.}\ \bibnamefont {Prilutskaya}}, \bibinfo
  {author} {\bibfnamefont {D.~A.}\ \bibnamefont {Skvortsova}}, \bibinfo
  {author} {\bibfnamefont {T.~R.}\ \bibnamefont {Zagirov}}, \bibinfo {author}
  {\bibfnamefont {D.~B.}\ \bibnamefont {Tretyakov}}, \bibinfo {author}
  {\bibfnamefont {V.~M.}\ \bibnamefont {Entin}}, \bibinfo {author}
  {\bibfnamefont {N.~N.}\ \bibnamefont {Bezuglov}},\ and\ \bibinfo {author}
  {\bibfnamefont {I.~I.}\ \bibnamefont {Ryabtsev}},\ }\bibfield  {title}
  {\bibinfo {title} {Rabi oscillations at three-photon laser excitation of a
  single rubidium rydberg atom in an optical dipole trap},\ }\href
  {https://doi.org/10.31857/S0044451024100109} {\bibfield  {journal} {\bibinfo
  {journal} {Zh. Eksp. Teor. Fiz.}\ }\textbf {\bibinfo {volume} {166}},\
  \bibinfo {pages} {535} (\bibinfo {year} {2024})},\ \bibinfo {note} {[English
  translation at arXiv: \url{https://arxiv.org/abs/2410.01703}]}\BibitemShut
  {NoStop}%
\bibitem [{\citenamefont {Grimm}\ \emph {et~al.}(2000)\citenamefont {Grimm},
  \citenamefont {Weidem\"{u}ller},\ and\ \citenamefont
  {Ovchinnikov}}]{Grimm2000}%
  \BibitemOpen
  \bibfield  {author} {\bibinfo {author} {\bibfnamefont {R.}~\bibnamefont
  {Grimm}}, \bibinfo {author} {\bibfnamefont {M.}~\bibnamefont
  {Weidem\"{u}ller}},\ and\ \bibinfo {author} {\bibfnamefont {Y.~B.}\
  \bibnamefont {Ovchinnikov}},\ }\bibinfo {title} {Optical dipole traps for
  neutral atoms},\ in\ \href {https://doi.org/10.1016/s1049-250x(08)60186-x}
  {\emph {\bibinfo {booktitle} {Advances In Atomic, Molecular, and Optical
  Physics}}}\ (\bibinfo  {publisher} {Elsevier},\ \bibinfo {year} {2000})\ p.\
  \bibinfo {pages} {95–170}\BibitemShut {NoStop}%
\bibitem [{\citenamefont {Shore}(2009)}]{Shore2009}%
  \BibitemOpen
  \bibfield  {author} {\bibinfo {author} {\bibfnamefont {B.}~\bibnamefont
  {Shore}},\ }\href {https://doi.org/10.1017/cbo9780511675713} {\emph {\bibinfo
  {title} {Manipulating Quantum Structures Using Laser Pulses}}}\ (\bibinfo
  {publisher} {Cambridge University Press},\ \bibinfo {year}
  {2009})\BibitemShut {NoStop}%
\bibitem [{\citenamefont {Cohen-Tannoudji}(1998)}]{CohenTannoudji1998}%
  \BibitemOpen
  \bibfield  {author} {\bibinfo {author} {\bibfnamefont {C.~N.}\ \bibnamefont
  {Cohen-Tannoudji}},\ }\bibfield  {title} {\bibinfo {title} {Nobel lecture:
  Manipulating atoms with photons},\ }\href
  {https://doi.org/10.1103/RevModPhys.70.707} {\bibfield  {journal} {\bibinfo
  {journal} {Rev. Mod. Phys.}\ }\textbf {\bibinfo {volume} {70}},\ \bibinfo
  {pages} {707} (\bibinfo {year} {1998})}\BibitemShut {NoStop}%
\bibitem [{\citenamefont {Stenholm}(2012)}]{stenholm2012foundations}%
  \BibitemOpen
  \bibfield  {author} {\bibinfo {author} {\bibfnamefont {S.}~\bibnamefont
  {Stenholm}},\ }\href {https://books.google.ru/books?id=ZKhXHSkTSbEC} {\emph
  {\bibinfo {title} {Foundations of Laser Spectroscopy}}},\ Dover Books on
  Physics\ (\bibinfo  {publisher} {Dover Publications},\ \bibinfo {year}
  {2012})\BibitemShut {NoStop}%
\bibitem [{\citenamefont {Macrì}\ \emph {et~al.}(2023)\citenamefont {Macrì},
  \citenamefont {Giannelli}, \citenamefont {Paladino},\ and\ \citenamefont
  {Falci}}]{Macr2023}%
  \BibitemOpen
  \bibfield  {author} {\bibinfo {author} {\bibfnamefont {N.}~\bibnamefont
  {Macrì}}, \bibinfo {author} {\bibfnamefont {L.}~\bibnamefont {Giannelli}},
  \bibinfo {author} {\bibfnamefont {E.}~\bibnamefont {Paladino}},\ and\
  \bibinfo {author} {\bibfnamefont {G.}~\bibnamefont {Falci}},\ }\bibfield
  {title} {\bibinfo {title} {Coarse-grained effective hamiltonian via the
  magnus expansion for a three-level system},\ }\bibfield  {journal} {\bibinfo
  {journal} {Entropy}\ }\textbf {\bibinfo {volume} {25}},\ \href
  {https://doi.org/10.3390/e25020234} {10.3390/e25020234} (\bibinfo {year}
  {2023})\BibitemShut {NoStop}%
\bibitem [{\citenamefont {Kirova}\ \emph {et~al.}(2017)\citenamefont {Kirova},
  \citenamefont {Cinins}, \citenamefont {Efimov}, \citenamefont {Bruvelis},
  \citenamefont {Miculis}, \citenamefont {Bezuglov}, \citenamefont {Auzinsh},
  \citenamefont {Ryabtsev},\ and\ \citenamefont {Ekers}}]{Kirova2017}%
  \BibitemOpen
  \bibfield  {author} {\bibinfo {author} {\bibfnamefont {T.}~\bibnamefont
  {Kirova}}, \bibinfo {author} {\bibfnamefont {A.}~\bibnamefont {Cinins}},
  \bibinfo {author} {\bibfnamefont {D.~K.}\ \bibnamefont {Efimov}}, \bibinfo
  {author} {\bibfnamefont {M.}~\bibnamefont {Bruvelis}}, \bibinfo {author}
  {\bibfnamefont {K.}~\bibnamefont {Miculis}}, \bibinfo {author} {\bibfnamefont
  {N.~N.}\ \bibnamefont {Bezuglov}}, \bibinfo {author} {\bibfnamefont
  {M.}~\bibnamefont {Auzinsh}}, \bibinfo {author} {\bibfnamefont {I.~I.}\
  \bibnamefont {Ryabtsev}},\ and\ \bibinfo {author} {\bibfnamefont
  {A.}~\bibnamefont {Ekers}},\ }\bibfield  {title} {\bibinfo {title} {Hyperfine
  interaction in the autler-townes effect: The formation of bright, dark, and
  chameleon states},\ }\href {https://doi.org/10.1103/PhysRevA.96.043421}
  {\bibfield  {journal} {\bibinfo  {journal} {Phys. Rev. A}\ }\textbf {\bibinfo
  {volume} {96}},\ \bibinfo {pages} {043421} (\bibinfo {year}
  {2017})}\BibitemShut {NoStop}%
\bibitem [{\citenamefont {Cinins}\ \emph {et~al.}(2024)\citenamefont {Cinins},
  \citenamefont {Efimov}, \citenamefont {Bruvelis}, \citenamefont {Miculis},
  \citenamefont {Kirova}, \citenamefont {Bezuglov}, \citenamefont {Ryabtsev},
  \citenamefont {Auzinsh},\ and\ \citenamefont {Ekers}}]{PhysRevA.109.063116}%
  \BibitemOpen
  \bibfield  {author} {\bibinfo {author} {\bibfnamefont {A.}~\bibnamefont
  {Cinins}}, \bibinfo {author} {\bibfnamefont {D.~K.}\ \bibnamefont {Efimov}},
  \bibinfo {author} {\bibfnamefont {M.}~\bibnamefont {Bruvelis}}, \bibinfo
  {author} {\bibfnamefont {K.}~\bibnamefont {Miculis}}, \bibinfo {author}
  {\bibfnamefont {T.}~\bibnamefont {Kirova}}, \bibinfo {author} {\bibfnamefont
  {N.~N.}\ \bibnamefont {Bezuglov}}, \bibinfo {author} {\bibfnamefont {I.~I.}\
  \bibnamefont {Ryabtsev}}, \bibinfo {author} {\bibfnamefont {M.}~\bibnamefont
  {Auzinsh}},\ and\ \bibinfo {author} {\bibfnamefont {A.}~\bibnamefont
  {Ekers}},\ }\bibfield  {title} {\bibinfo {title} {Hyperfine interaction in
  the autler-townes effect: Control of two-photon selection rules in the
  morris-shore basis},\ }\href {https://doi.org/10.1103/PhysRevA.109.063116}
  {\bibfield  {journal} {\bibinfo  {journal} {Phys. Rev. A}\ }\textbf {\bibinfo
  {volume} {109}},\ \bibinfo {pages} {063116} (\bibinfo {year}
  {2024})}\BibitemShut {NoStop}%
\bibitem [{\citenamefont {Kazansky}\ \emph {et~al.}(2001)\citenamefont
  {Kazansky}, \citenamefont {Bezuglov}, \citenamefont {Molisch}, \citenamefont
  {Fuso},\ and\ \citenamefont {Allegrini}}]{Kazansky}%
  \BibitemOpen
  \bibfield  {author} {\bibinfo {author} {\bibfnamefont {A.~K.}\ \bibnamefont
  {Kazansky}}, \bibinfo {author} {\bibfnamefont {N.~N.}\ \bibnamefont
  {Bezuglov}}, \bibinfo {author} {\bibfnamefont {A.~F.}\ \bibnamefont
  {Molisch}}, \bibinfo {author} {\bibfnamefont {F.}~\bibnamefont {Fuso}},\ and\
  \bibinfo {author} {\bibfnamefont {M.}~\bibnamefont {Allegrini}},\ }\bibfield
  {title} {\bibinfo {title} {Direct numerical method to solve radiation
  trapping problems with a doppler-broadening mechanism for partial frequency
  redistribution},\ }\href {https://doi.org/10.1103/PhysRevA.64.022719}
  {\bibfield  {journal} {\bibinfo  {journal} {Phys. Rev. A}\ }\textbf {\bibinfo
  {volume} {64}},\ \bibinfo {pages} {022719} (\bibinfo {year}
  {2001})}\BibitemShut {NoStop}%
\bibitem [{\citenamefont {Efimov}\ \emph {et~al.}(2014)\citenamefont {Efimov},
  \citenamefont {Bezuglov}, \citenamefont {Klyucharev}, \citenamefont {Gnedin},
  \citenamefont {Miculis},\ and\ \citenamefont {Ekers}}]{efimov2014}%
  \BibitemOpen
  \bibfield  {author} {\bibinfo {author} {\bibfnamefont {D.~K.}\ \bibnamefont
  {Efimov}}, \bibinfo {author} {\bibfnamefont {N.~N.}\ \bibnamefont
  {Bezuglov}}, \bibinfo {author} {\bibfnamefont {A.~N.}\ \bibnamefont
  {Klyucharev}}, \bibinfo {author} {\bibfnamefont {Y.~N.}\ \bibnamefont
  {Gnedin}}, \bibinfo {author} {\bibfnamefont {K.}~\bibnamefont {Miculis}},\
  and\ \bibinfo {author} {\bibfnamefont {A.}~\bibnamefont {Ekers}},\ }\bibfield
   {title} {\bibinfo {title} {Analysis of light-induced diffusion ionization of
  a three-dimensional hydrogen atom based on the floquet technique and
  split-operator method},\ }\href {https://doi.org/10.1134/S0030400X1407008X}
  {\bibfield  {journal} {\bibinfo  {journal} {Optics and Spectroscopy}\
  }\textbf {\bibinfo {volume} {117}},\ \bibinfo {pages} {8} (\bibinfo {year}
  {2014})}\BibitemShut {NoStop}%
\bibitem [{\citenamefont {Hairer}\ \emph {et~al.}(2006)\citenamefont {Hairer},
  \citenamefont {Wanner},\ and\ \citenamefont {Lubich}}]{Hairer2006}%
  \BibitemOpen
  \bibfield  {author} {\bibinfo {author} {\bibfnamefont {E.}~\bibnamefont
  {Hairer}}, \bibinfo {author} {\bibfnamefont {G.}~\bibnamefont {Wanner}},\
  and\ \bibinfo {author} {\bibfnamefont {C.}~\bibnamefont {Lubich}},\ }\href
  {https://doi.org/10.1007/3-540-30666-8} {\emph {\bibinfo {title} {Geometric
  Numerical Integration: Structure-Preserving Algorithms for Ordinary
  Differential Equations}}}\ (\bibinfo  {publisher} {Springer Berlin
  Heidelberg},\ \bibinfo {address} {Berlin, Heidelberg},\ \bibinfo {year}
  {2006})\BibitemShut {NoStop}%
\bibitem [{\citenamefont {Hairer}\ \emph {et~al.}(2003)\citenamefont {Hairer},
  \citenamefont {Lubich},\ and\ \citenamefont {Wanner}}]{Hairer2003}%
  \BibitemOpen
  \bibfield  {author} {\bibinfo {author} {\bibfnamefont {E.}~\bibnamefont
  {Hairer}}, \bibinfo {author} {\bibfnamefont {C.}~\bibnamefont {Lubich}},\
  and\ \bibinfo {author} {\bibfnamefont {G.}~\bibnamefont {Wanner}},\
  }\bibfield  {title} {\bibinfo {title} {Geometric numerical integration
  illustrated by the st\"{o}rmer–verlet method},\ }\href
  {https://doi.org/10.1017/S0962492902000144} {\bibfield  {journal} {\bibinfo
  {journal} {Acta Numerica}\ }\textbf {\bibinfo {volume} {12}},\ \bibinfo
  {pages} {399–450} (\bibinfo {year} {2003})}\BibitemShut {NoStop}%
\bibitem [{\citenamefont {Sydoryk}\ \emph {et~al.}(2008)\citenamefont
  {Sydoryk}, \citenamefont {Bezuglov}, \citenamefont {Beterov}, \citenamefont
  {Miculis}, \citenamefont {Saks}, \citenamefont {Janovs}, \citenamefont
  {Spels},\ and\ \citenamefont {Ekers}}]{PhysRevA.77.042511}%
  \BibitemOpen
  \bibfield  {author} {\bibinfo {author} {\bibfnamefont {I.}~\bibnamefont
  {Sydoryk}}, \bibinfo {author} {\bibfnamefont {N.~N.}\ \bibnamefont
  {Bezuglov}}, \bibinfo {author} {\bibfnamefont {I.~I.}\ \bibnamefont
  {Beterov}}, \bibinfo {author} {\bibfnamefont {K.}~\bibnamefont {Miculis}},
  \bibinfo {author} {\bibfnamefont {E.}~\bibnamefont {Saks}}, \bibinfo {author}
  {\bibfnamefont {A.}~\bibnamefont {Janovs}}, \bibinfo {author} {\bibfnamefont
  {P.}~\bibnamefont {Spels}},\ and\ \bibinfo {author} {\bibfnamefont
  {A.}~\bibnamefont {Ekers}},\ }\bibfield  {title} {\bibinfo {title}
  {Broadening and intensity redistribution in the $\text{Na}(3p)$ hyperfine
  excitation spectra due to optical pumping in the weak excitation limit},\
  }\href {https://doi.org/10.1103/PhysRevA.77.042511} {\bibfield  {journal}
  {\bibinfo  {journal} {Phys. Rev. A}\ }\textbf {\bibinfo {volume} {77}},\
  \bibinfo {pages} {042511} (\bibinfo {year} {2008})}\BibitemShut {NoStop}%
\end{thebibliography}%
\end{document}